\documentclass[prd,preprint,superscriptaddress,nofootinbib]{revtex4}

\usepackage[pdftex]{graphicx,color}
\usepackage{epstopdf}
\usepackage[bookmarks=true,bookmarksnumbered=true,bookmarkstype=toc]{hyperref} 
\usepackage{units}

\usepackage{graphicx}
\usepackage{color}
\usepackage{amsfonts}
\usepackage{latexsym}
\usepackage{amsmath}
\usepackage{amssymb}
\usepackage{epsfig}
\usepackage{enumitem}
\usepackage{slashed}
\usepackage{physics2}
\usepackage{tikz-feynhand}
\makeatletter
\let\MYcaption\@makecaption
\makeatother

\usepackage{subcaption}
\makeatletter
\let\@makecaption\MYcaption
\makeatother
\usepackage{color}

\definecolor{green}{cmyk}{1,0,1,0}

\definecolor{pink}{cmyk}{0,0.5,0,0}

\definecolor{pastelpink}{cmyk}{0,0.25,0,0}

\definecolor{softpink}{cmyk}{0,0.125,0,0}

\definecolor{purple}{cmyk}{0.5,1.0,0.1,0}

\definecolor{violet}{cmyk}{0.75,1,0.25,0}

\begin{document}

\title{
New rare meson decay constraints on a light vector in $U(1)_{B-L}, U(1)_R$ and the dark photon 
}

\author{Osamu Seto}
 \email{seto@particle.sci.hokudai.ac.jp}
 \affiliation{Department of Physics, Hokkaido University, Sapporo 060-0810, Japan}

\author{Takashi Shimomura}
\email{shimomura@cc.miyazaki-u.ac.jp}
\affiliation{Faculty of Education, Miyazaki University, Miyazaki 889-2192, Japan}

\author{Shinsuke Yoshida}
 \affiliation{Department of Physics, Hokkaido University, Sapporo 060-0810, Japan}

%%%%%%%%%%%%%%%%%%%%%%
\begin{abstract}
%%%%%%%%%%%%%%%%%%%%%%
We evaluate constraints from flavor changing rare meson decays to a light vector boson $X$, followed by the decay of the on-shell $X$ into the SM fermions. The flavor changing meson decay emitting the light $X$ is induced by loop processes where the up-type quarks, the $W$ boson, or charged scalar bosons are running inside loops. We calculate all one-loop diagrams with neglecting all masses of light quarks except for the top quark in a general anomaly free extra $U(1)$ model. Our theoretical evaluation of the branching ratio of charged $B$ meson decay and charged kaon decay is compared to experimental results, and we derive new constraints for dark photon,  $U(1)_{B-L}$ and $U(1)_R$ models. 
%%%%%%%%%%%%%%%%%%%%%%
\end{abstract}
%%%%%%%%%%%%%%%%%%%%%%

%\pacs{}
\preprint{EPHOU-25-006} 
\preprint{UME-PP-029}

\vspace*{3cm}

\maketitle
%==================================%
%          Main body               %
%==================================%

%%%%%%%%%%%%%%%%%%%%%%%
\section{Introduction}
%%%%%%%%%%%%%%%%%%%%%%%

Recently, extensions of the standard model (SM) of particle physics
 with feebly interacting light particles have been subjects of
 interest~\cite{Bauer:2018onh,Fabbrichesi:2020wbt,Lanfranchi:2020crw,Caputo:2021eaa}.
Those particles often appear in interesting particle models of beyond the SM (BSM).
The global $B-L$ (baryon number$-$lepton number) symmetry or the right-handed chirality
 would be gauged as $U(1)_{B-L}$~\cite{Pati:1973uk,Davidson:1978pm,Mohapatra:1980qe,Mohapatra:1980} or $U(1)_R$~\cite{Jung:2009jz} as an anomaly free renormalizable model. 
Those models are well-motivated from the viewpoint of neutrino mass generation~\cite{Mohapatra:1980qe} and
 the new gauge interaction also makes sterile neutrino dark matter viable~\cite{Khalil:2008kp,Kaneta:2016vkq,Eijima:2022dec,Seto:2024lik}.
If its gauge coupling is very small, the gauge boson $X$ is a feeble light particle. 
A simpler phenomenological example is so-called dark photon $A'$ where a Proca field couples with the photon (or the $U(1)_Y$ gauge field) only through the tiny kinetic mixing~\cite{Galison:1983pa,Holdom:1985ag}.
Even if the coupling constant is very small, it would be possible for any feebly interacting particle
 to be produced if its mass is small enough, 
 since various present experiments with huge integrated luminosity have been in progress. 

Direct searches of a feeble particle through its decay into visible particles have been carried out in 
 collider and fixed target experiments.
Null results in collider experiments such as CMS~\cite{CMS:2019buh}, LHCb~\cite{LHCb:2019vmc}, Babar~\cite{BaBar:2014zli,BaBar:2017tiz} and NA48~\cite{NA482:2015wmo} so far exclude coupling constants larger than $\mathcal{O}(10^{-3})$.
Since a feeble particle is long-lived due to its small coupling and leaves the displaced vertex,
 those can be searched as long-lived particles (LLPs).
Fixed target and beam dump experiments such as NuCal~\cite{Blumlein:2013cua}, E137~\cite{Marsicano:2018krp} and CHARM~\cite{Gninenko:2012eq} have better sensitivity than collider experiments for a smaller coupling constant for the sake of distant detectors. 
Since the sensitivity of those experiments are limited by mainly the beam energy, luminosity,
 detector position from colliding points, combination of collider experiments
 and fixed target experiments are not only complementary but also leave unexplored parameter region.

In proton beam dump and fixed target experiments,
 a feeble light vector boson $X$ is mainly produced by the decay of SM mesons as well as Bremsstrahlung processes.
Hence, the interaction between $X$ and quarks in a meson plays an important role for the production.
By the same token, for a decay mode of a meson $M$ to another meson $M'$ emitting $X$,
 such flavor changing meson decay are well measured by many experiments and consistent with the SM.
Couplings between $X$ and quarks are constrained by even little deviation in meson decay.

Through studies on various rare meson decay modes, principally the decay of $B_s$ meson and other modes~\cite{Davoudiasl:2012ag,Xu:2015wja,Datta:2022zng},
 it turns out that the mixing between the $Z$ boson and $X$ in the mass term has to be very small~\cite{Datta:2022zng}.
The dark photon effect on the decay mode $K_L \rightarrow \pi^0 \nu\bar{\nu}$ is less than a few percent to the SM contribution~\cite{Wang:2023css}. 
Dror et al have calculated, from the effective field theory (EFT) perspective, the bounds from various flavor changing meson decay modes emitting $X$ by utilizing the Goldstone boson equivalence in Refs.~\cite{Dror:2017ehi,Dror:2017nsg}.
On the other hand, in this paper, we perform straightforward standard calculation to derive loop induced flavor changing couplings in an ultraviolet (UV) renormalizable anomaly free models based on an extra $U(1)_X$ gauge symmetry.  
The extra $U(1)_X$ gauge symmetry with $X$ is expressed as a linear combination of $U(1)_Y$ and $U(1)_{B-L}$.
We apply our results on the minimal $U(1)_{B-L}$ model, the minimal $U(1)_R$ model and the minimal dark photon model.
We find that wide parameter range of unexplored region by direct search experiments has been already, in fact, constrained by the flavor changing rare meson decay for all three models.

This paper is organized as follows. 
In the next section, we introduce models for an extra $U(1)$ gauge symmetry,
and note formula of mass eigenstates and relevant interaction vertexes, which are mostly in Appendices.
In Sec.~\ref{sec:FCmesondecay}, we present formulae to estimate a flavor changing meson decay.
We investigate three specific models in Sec.~\ref{sec:Results}. 
Section~\ref{sec:Summary} is devoted to summary.

%%%%%%%%%%%%%%%%%%%%%%%
\section{Model}
%%%%%%%%%%%%%%%%%%%%%%%

\subsection{$U(1)_X$ model}

%%%%%%%%%%%%%%%%%%%%%%%%%%%%%%%%%%%%%%
\begin{table}[htbp]
	\centering
	\begin{tabular}{|c|ccc|c|} \hline	
		 & $SU(3)_C$ & $SU(2)_L$ & $U(1)_Y$  & $U(1)_X $ \\ \hline
		$Q_L^i$ & $\mathbf{3}$ & $\mathbf{2}$ & $\frac{1}{6}$ & $\frac{1}{6}x_H+\frac{1}{3}$ \\
		$u_R^i$ & $\mathbf{3}$ & $\mathbf{1}$ & $\frac{2}{3}$ & $\frac{2}{3}x_H+\frac{1}{3}$ \\
		$d_R^i$ & $\mathbf{3}$ & $\mathbf{1}$ & $-\frac{1}{3}$ & $-\frac{1}{3}x_H+\frac{1}{3}$ \\ \hline
		$L_L^i$ & $\mathbf{1}$ & $\mathbf{2}$ & $-\frac{1}{2}$ & $-\frac{1}{2}x_H-1$ \\
		$e_R^i$ & $\mathbf{1}$ & $\mathbf{1}$ & $-1$ & $-x_H-1$ \\
		$\nu_R^i$ & $\mathbf{1}$ & $\mathbf{1}$ & $0$ & $-1$ \\  \hline
        $\Phi_1$ & $\mathbf{1}$ & $\mathbf{2}$ & $\frac{1}{2}$ & $\frac{1}{2}x_H+x_{\Phi}$ \\
        $\Phi_2$ & $\mathbf{1}$ & $\mathbf{2}$ & $\frac{1}{2}$ & $\frac{1}{2}x_H$ \\ 
        $\Phi_X$ & $\mathbf{1}$ & $\mathbf{1}$ & $0$ & $2$ \\ \hline
	\end{tabular}
\caption{
In addition to the SM particle content ($i=1,2,3$), three right-handed neutrinos  
 $\nu_R^i$ ($i=1, 2, 3$), two Higgs doublet fields $\Phi_1$ and $\Phi_2$,
 and one $U(1)_X$ Higgs field $\Phi_X$ are introduced. 
$x_H$ is a real free parameter in the $U(1)_X$ charge unfixed by the anomaly-free conditions.
}
\label{table1}
\end{table}
%%%%%%%%%%%%%%%%%%%%%%%%%%%%%%%%%%%%%%%%%%%%%%%

We consider the model based on the gauge group $SU(3)_C \times SU(2)_L \times U(1)_Y \times U(1)_X$, 
where $U(1)_X$ is a linear combination of $U(1)_Y$ and $U(1)_{B-L}$~\cite{Appelquist:2002mw}. 
A parameter $x_H$ parameterizes the relative $U(1)_Y$ charge with respect to the $U(1)_{B-L}$ charge in the $U(1)_X$ charge.
The case with $x_H =0$ corresponds to $U(1)_{B-L}$ and that with $x_H=-2$ does to $U(1)_R$.
The $U(1)_X$ charge is listed in Tab.~\ref{table1}.
We introduce a singlet Higgs field $\Phi_X$ responsible to break the $U(1)_X$ gauge symmetry.
Our model contains two Higgs doublet fields $\Phi_1$ and $\Phi_2$, and those are charged under the $U(1)_X$ with
 the charge $x_H/2+x_{\Phi}$ and $x_H/2$, respectively.
$\Phi_2$ forms Yukawa couplings with the SM fermions and generates those masses even for $x_H \neq 0$~\cite{Oda:2015gna,Das:2016zue}.
$\Phi_1$ is needed to cancel the mass mixing between the $Z$ and $X$ bosons and
 must have the different $U(1)_X$ charge from $\Phi_2$ to remove an unwanted Nambu-Goldstone (NG) mode~\cite{Seto:2020jal}. 

The Yukawa couplings are given by
\begin{align}
 \mathcal{L}_\mathrm{Yukawa} = 
  & -y^{u} \overline{Q} \tilde{\Phi}_2 u_R -y^{d} \overline{Q} \Phi_2 d_R -y^{e} \overline{L} \Phi_2 e_R
    -y^{\nu} \overline{L} \tilde{\Phi}_2 \nu_R 
    - \frac{1}{2}y^{\nu_R} \overline{\nu_R^{C}} \Phi_X \nu_R + \mathrm{ H.c.} , \\
    \tilde{\Phi} =& i \tau_2 \Phi^*,
\label{Lag:yukawa} 
\end{align}
 where the superscript $C$ denotes the charge conjugation, and the generation indexes are suppressed.

The gauge kinetic terms are
\begin{align}
\mathcal{L}_{\mathrm{Gauge}} =& -\frac{1}{4}\hat{W}_{\mu\nu}\hat{W}^{\mu\nu} -\frac{1}{4}\hat{B}_{\mu\nu}\hat{B}^{\mu\nu} -\frac{1}{4} \hat{X}_{\mu\nu}\hat{X}^{\mu\nu} +\frac{\sin\epsilon}{2}\hat{B}_{\mu\nu}\hat{X}^{\mu\nu},
\label{Lag:gauge} 
\end{align}
where $\hat{W}_{\mu\nu}, \hat{B}_{\mu\nu}$ and $\hat{X}_{\mu\nu}$ are the field strength of $SU(2)_L, U(1)_Y$ and $U(1)_X$, respectively. 
The hat stands for those in gauge eigenstates and $\epsilon$ is the gauge kinetic mixing parameter. The gauge field strength of $SU(3)_C$ interaction is omitted.

The Higgs part is given by
\begin{align}
 \mathcal{L}_\mathrm{Higgs} &= |D_{\mu}\Phi_1|^2 +|D_{\mu}\Phi_2|^2 + |D_{\mu}\Phi_X|^2 -V, \\
 D_{\mu}&= \partial_{\mu} - i\frac{1}{2}g_2 \tau\!\cdot\! \hat{W}_{\mu}- i g_1 q_Y \hat{B}_{\mu}- i g_X q_X \hat{X}_{\mu}, 
\label{Lag:higgs} 
\end{align}
where $g_2, g_1$, and $g_X$ are the gauge coupling constants for $SU(2)_L, U(1)_Y$ and $U(1)_X$,
 and $q_Y$ and $q_X$ are those charges for each Higgs field. 
 The $SU(2)_L$ generator is denoted as $\tau \equiv \sigma/2$, $\sigma$ being the Pauli matrix, and dot symbol represents 
 the inner product over $SU(2)_L$ index.
The scalar potential is given by
\begin{align}
V =& V_1 +V_2, \label{eq:V_UV} \\
V_1 =&  \hat{\mu_1}^2 |\Phi_1|^2 -\hat{\mu_2}^2 |\Phi_2|^2  + \frac{1}{2}\lambda_1|\Phi_1|^4   +
 \frac{1}{2}\lambda_2|\Phi_2|^4-\mu_X^2 |\Phi_X|^2 + \frac{1}{2}\lambda_X |\Phi_X|^4  \nonumber\\ 
 & + \lambda_3|\Phi_1|^2|\Phi_2|^2 +\lambda_4 |\Phi_1^\dagger \Phi_2|^2 + \lambda_5|\Phi_1|^2|\Phi_X|^2 +\lambda_6 |\Phi_2|^2|\Phi_X|^2 ,\\
V_2 = &
\left\{
\begin{array}{ccc}
 \lambda_{12}^{(1)} \Phi_X (\Phi_1^\dagger \Phi_2) + \mathrm{ H.c.}   &\quad & x_{\Phi}=2 \\
 \lambda_{12}^{(2)} \Phi_X (\Phi_1 \Phi_2^\dagger) + \mathrm{ H.c.}   && x_{\Phi}=-2 \\
 \lambda_{12}^{(3)} \Phi_X^2 (\Phi_1^\dagger \Phi_2) + \mathrm{ H.c.} && x_{\Phi}=4 \\
 \lambda_{12}^{(4)} \Phi_X^2 (\Phi_1 \Phi_2^\dagger) + \mathrm{ H.c.} && x_{\Phi}=-4 \\
\end{array}
\right. . \label{eq:V2} 
\end{align}
The $V_2$ part removes the unwanted NG mode~\cite{Seto:2020jal}.

\subsection{Mass eigenstates in the broken vacuum}

At the EW and $U(1)_X$ breaking vacuum, 
each Higgs fields are expanded around the vacuum expectation values (VEVs), $v_1, v_2$ and $v_X$, as 
\begin{align}
\Phi_1 & = \left( 
\begin{array}{c}
 \phi_1^+ \\
 \frac{1}{\sqrt{2}}(v_1 + \phi_1 + i a_1)  \\
\end{array} \right) , \\
\Phi_2 & = \left( 
\begin{array}{c}
 \phi_2^+ \\
 \frac{1}{\sqrt{2}}(v_2 + \phi_2 + i a_2)  \\
\end{array} \right) , \\
\Phi_X & =  \frac{1}{\sqrt{2}}(v_X + \phi_X + i a_X)  ,
\end{align}
 with $v=\sqrt{v_1^2+v_2^2}\simeq 246$ GeV. We define $\tan\beta=v_2/v_1$.
The mass eigenstates of CP even neutral Higgs bosons $h, H$ and $\varphi$
 are obtained through the unitary matrix $U_S$ as
\begin{align}
\left( 
\begin{array}{c}
\phi_1 \\
\phi_2 \\
\phi_X
\end{array}
\right)
= U_S
\left( 
\begin{array}{c}
h \\
H \\
\varphi
\end{array}
\right) ,
\label{eq:mixing:h}
\end{align}
 where $h$ is identified as the SM-like Higgs boson with the mass $m_h \simeq 125$ GeV,
 $\varphi$ is the singlet-like Higgs boson, and $H$ is the heavier Higgs boson. 
Similarly, the charged Higgs boson $H^+$, the CP odd Higgs boson $A$ and NG modes are obtained
 through the unitary matrix $U_C$ and $U_P$ as
\begin{align}
\left( 
\begin{array}{c}
\phi^+_1 \\
\phi^+_2 
\end{array}
\right)
= U_C
\left( 
\begin{array}{c} 
\omega^+ \\
H^+
\end{array}
\right) ,
\label{eq:mixing:H+}
\end{align}
\begin{align}
\left( 
\begin{array}{c}
a_1 \\
a_2 \\
a_X
\end{array}
\right)
= U_P
\left( 
\begin{array}{c} 
\hat{\omega}_z \\
\hat{\omega}_x \\
A
\end{array}
\right) ,
\label{eq:mixing:A}
\end{align}
where $\omega^+, \hat{\omega}_z$ and $\hat{\omega}_x$ are NG modes eaten by the gauge bosons respectively.
The expressions of the unitary matrices for $x_{\Phi}=2$, as an example, can be found in Appendix~\ref{sec:App:Higgs}.

We define
\begin{align}
\hat{A}_{\mu} &= s_W \hat{W}^3_{\mu} + c_W \hat{B}_{\mu} , \\
\hat{Z}_{\mu} &= c_W \hat{W}^3_{\mu} - s_W \hat{B}_{\mu} ,
\end{align}
with $s_W = \sin\theta_W = g_1/\sqrt{g_2^2+g_1^2}$ and $c_W = \cos\theta_W$, where $\theta_W$ is the Weinberg angle. Here and hereafter, we use similar abbreviation of trigonometric functions.
Now, we will find the transformation from the hatted states, $\hat{A}_{\mu}, \hat{Z}_{\mu}$ and$ \hat{X}_{\mu}$, to the mass eigenstates $A_{\mu}, Z_{\mu}$ and $ X_{\mu}$ by resolving the mixings.
In fact, the mixing between $\hat{Z}_{\mu}$ and $\hat{X}_{\mu}$ must be very small for the light $X$ boson to be consistent with the stringent bounds from neutrino-electron scattering measured at TEXONO~\cite{TEXONO:2009knm,TEXONO:2010tnr}.
The mass matrix of neutral gauge bosons is 
\begin{align}
& \left( 
\begin{array}{cc}
 M^2_{\hat{Z}}  & M^2_{\hat{Z}\hat{X}}\\
 M^2_{\hat{Z}\hat{X}} & M^2_{\hat{X}}\\
\end{array}\right)  \nonumber \\
 = & \left( 
\begin{array}{cc}
 \frac{g_1^2+g_2^2}{4} v^2 & -\frac{g_X \sqrt{g_1^2+g_2^2}}{4}\left(v_1^2 (2 x_{\Phi}+x_H)+v_2^2 x_H\right) \\
 -\frac{g_X \sqrt{g_1^2+g_2^2}}{4}\left(v_1^2 (2 x_{\Phi}+x_H)+v_2^2 x_H\right) & \frac{g_X^2}{4} \left(v_1^2 (2 x_{\Phi}+x_H)^2+v_2^2 x_H^2+16 v_X^2\right)
\end{array}\right) .
\end{align}
To suppress the mass mixing, we will take $\tan\beta$ appropriately for given $x_{\Phi}$ and $x_H$ so that the off-diagonal elements $M^2_{\hat{Z}\hat{X}}$, in other words the mass mixing, vanish. 
In fact, for this purpose, we have introduced the second Higgs doublet field $\Phi_1$.
The field redefinition by an orthogonal matrix, 
\begin{equation}
U_K = \left(
\begin{array}{ccc}
1 & 0 & t_{\epsilon} c_W \\
0 & 1 & -t_{\epsilon} s_W \\
0 & 0 & \frac{1}{c_{\epsilon}} \\
\end{array}
\right) ,
\end{equation}
 resolves the kinetic mixing and leaves the mass mixing 
\begin{equation}
M_V^2 = \left(
\begin{array}{ccc}
 0 & 0 & 0 \\
 0 & M^2_{\hat{Z}} & \frac{M^2_{\hat{Z}\hat{X}}}{c_{\epsilon}}-M^2_{\hat{Z}} s_W t_{\epsilon} \\
 0 & \frac{M^2_{\hat{Z}\hat{X}}}{c_{\epsilon}}-M^2_{\hat{Z}} s_W t_{\epsilon} & \frac{1}{c_{\epsilon}^2} \left(  s_W c_{\epsilon}t_{\epsilon} \left(  M^2_{\hat{Z}}s_W  c_{\epsilon} t_{\epsilon}-2 M^2_{\hat{Z}\hat{X}} \right)+M^2_{\hat{X}} \right) \\
\end{array}
\right) . \label{eq:MV2}
\end{equation}
The additional field redefinition to the mass eigenstates can be done with a rotation matrix
\begin{equation}
U_M = \left(
\begin{array}{ccc}
1 & 0 & 0 \\
0 & \cos\theta & -\sin\theta \\
0 & \sin\theta & \cos\theta \\
\end{array}
\right) ,
\end{equation}
 with the angle
\begin{equation}
\tan 2\theta = \frac{2 M_V^2{}_{23}}{M_V^2{}_{22}-M_V^2{}_{33} } ,
\label{eq:tan2theta}
\end{equation}
 where $M_V^2{}_{ij}$ are an element in Eq.~(\ref{eq:MV2}). Equation (\ref{eq:tan2theta}) is simplified as
\begin{equation}
 \tan 2\theta \simeq -2s_W t_{\epsilon} , \label{eq:theta-approx}
\end{equation}
for $\epsilon \ll 1$, $M^2_{\hat{X}} \ll M^2_{\hat{Z}} $ and $M^2_{\hat{Z}\hat{X}}=0$.
The mass eigenvalues of the mass eigenstates $Z$ and $X$ are given by
\begin{align}
m_Z^2 &= M_V^2{}_{22} \cos^2(\theta)+M_V^2{}_{23} \sin (2\theta)+M_V^2{}_{33} \sin^2(\theta), \\
m_X^2 &= M_V^2{}_{22} \sin^2(\theta)-M_V^2{}_{23} \sin (2\theta)+M_V^2{}_{33} \cos^2(\theta) .
\end{align}
Since the hatted field and the unhatted field of mass eigenstates are related as
\begin{align} 
\left(
\begin{array}{c}
\hat{A}_{\mu} \\
\hat{Z}_{\mu} \\
\hat{X}_{\mu} \\
\end{array}
\right) = U_K U_M
\left(
\begin{array}{c}
A_{\mu} \\
Z_{\mu} \\
Z'_{\mu} \\
\end{array}
\right) 
=
\left(
\begin{array}{ccc}
 1 & c_W s_{\theta} t_{\epsilon} & c_{\theta} c_W t_{\epsilon} \\
 0 & c_{\theta}-s_{\theta} s_W t_{\epsilon} & -c_{\theta} s_W t_{\epsilon}-s_{\theta} \\
 0 & \frac{s_{\theta}}{c_{\epsilon}} & \frac{c_{\theta}}{c_{\epsilon}} \\
\end{array}
\right)
\left(
\begin{array}{c}
A_{\mu} \\
Z_{\mu} \\
X_{\mu} \\
\end{array}
\right) 
,
\end{align}
by combining with
\begin{align}
\left(
\begin{array}{c}
\hat{W}^3_{\mu} \\
\hat{B}_{\mu} \\
\hat{X}_{\mu} \\
\end{array}
\right) &= \left(
\begin{array}{ccc}
s_W & c_W & 0  \\
c_W & -s_W  & 0  \\
0 & 0 & 1 \\
\end{array}
\right)  
\left(
\begin{array}{c}
\hat{A}_{\mu} \\
\hat{Z}_{\mu} \\
\hat{X}_{\mu} \\
\end{array}
\right) ,
\end{align}
 we find
\begin{align}
\left(
\begin{array}{c}
\hat{W}^3_{\mu} \\
\hat{B}_{\mu} \\
\hat{X}_{\mu} \\
\end{array}
\right) 
 = \left(
\begin{array}{ccc}
s_W & c_W c_{\theta} & - c_W s_{\theta}  \\
c_W & -s_W c_{\theta} + t_{\epsilon}s_{\theta}  & s_W s_{\theta} + t_{\epsilon} c_{\theta}  \\
0 & \frac{1}{c_{\epsilon}}s_{\theta} & \frac{1}{c_{\epsilon}} c_{\theta} \\
\end{array}
\right)  
\left(
\begin{array}{c}
A_{\mu} \\
Z_{\mu} \\
X_{\mu} \\
\end{array}
\right) .
\label{eq:mixing:gauge}
\end{align}

By substituting Eqs.~(\ref{eq:mixing:h}), (\ref{eq:mixing:H+}), (\ref{eq:mixing:A}) and (\ref{eq:mixing:gauge}), into the Lagrangian,
 we obtain the interaction vertexes for the mass eigenstates.
We have listed relevant couplings for this study in Appendix~\ref{sec:App:vertex}.
For an operator $O_i$ containing $P_L$ or $P_R$, we define the coupling coefficients $C_{iL/R}$ as  
\begin{equation}
\mathcal{L} \supset C_{iL/R}  O_{iL/R} ,
\end{equation}
with $P_{L/R}$ being the chirality projection operator.

We remark that the minimal $U(1)_{B-L}$ model is reproduced for $x_H=0$ with taking the limit of
 $v_1 \rightarrow 0$ and $\hat{\mu}_1^2 \rightarrow \infty$ simultaneously so that $\Phi_1$ decouples.
In addition, by taking $g_X \rightarrow 0$ limit with keeping $m_X \simeq 2g_X v_X$ finite\footnote{See e.g, Refs.~\cite{Araki:2020wkq,Araki:2024uad} for models where dark photon mass is generated through the spontaneous symmetry breaking by the dark Higgs field.},
 our model reduces to so-called the dark photon model where
 the dark photon $A'$ couples with the electromagnetic current $J_{\mathrm{em}}$ as
\begin{align}
\mathcal{L} \supset  -e\varepsilon J_{\mathrm{em}}^{\mu}A'_{\mu},
\end{align}
 with $\varepsilon = c_W \epsilon$ and $e=g_2s_W$.

%%%%%%%%%%%%%%%%%%%%%%%
\section{$X$ mediated flavor changing meson decays}
\label{sec:FCmesondecay}
%%%%%%%%%%%%%%%%%%%%%%%

We will examine decay modes
\begin{align}
\mathrm{Br}\left( B^- \rightarrow K^- (X \rightarrow \ell^+\ell^-) \right)
= &\mathrm{Br}(B^- \rightarrow K^- X) \mathrm{Br}(X \rightarrow \ell^+\ell^-) ,  \label{eq:br-B}\\
\mathrm{Br}\left( K^- \rightarrow \pi^- (X \rightarrow \nu\bar{\nu}) \right)
= &\mathrm{Br}(K^- \rightarrow \pi^- X) \mathrm{Br}(X \rightarrow \nu\bar{\nu})  \label{eq:br-K},
\end{align}
where favor changing rare meson decay processes produce the $X$ boson on-shell, followed by the decay of $X$ into the SM fermions.
Since we consider on-shell production of $X$, 
to evaluate a flavor changing meson decay through the $X$ boson emission, 
we can separately calculate the $X$ decay and the $X$ production through meson decay.
On the other hand, the latest experimental result on $B^- \rightarrow K^-\ell^+\ell^-$ is applicable to 
the decay mode $ B^- \rightarrow K^- (X \rightarrow \ell^+\ell^-)$ if the on-shell $X$ decays promptly.
We impose the condition for the prompt decay of $X$ to the travel distance of $l_X$ as
\begin{equation}
l_X < 50\, \mu\mathrm{m} ,
\label{eq:prompt_decay}
\end{equation}
due to the spatial resolution of Belle II~\cite{Belle-II:2018jsg,Aihara:2024zds}.

\subsection{$X$ boson decays}

\begin{figure}[htbp]
\centering
\includegraphics[clip,width=12.0cm]{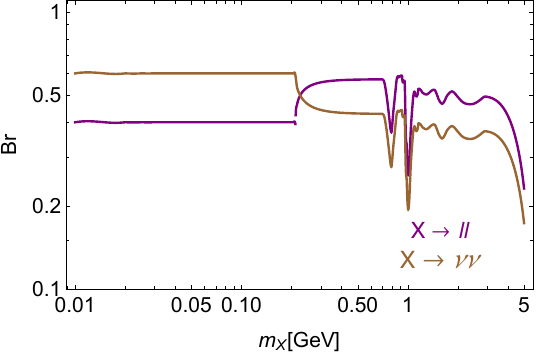}
\caption{
$\mathrm{Br}(X \rightarrow \ell^-\ell^+)$ (purple curve) and $\mathrm{Br}(X \rightarrow \nu\bar{\nu})$ (brown curve) 
for $x_H=0$ as a function of $m_X$. 
}
\label{Fig:XdecayBr}
\end{figure}

We first calculate the decay rate of the $X$ boson.
The partial decay rate of the $X$ boson into a pair of fermion $f=\nu_i, l_i, u_i, d_i$ with the mass $m_f$ and the corresponding decay branching ratio are given by
\begin{align}
\Gamma(X \rightarrow f\bar{f}) =& N_c\frac{\sqrt{m_X^2-4 m_f^2}}{24 \pi m_X^2}\left( \left(C_{fXL}^2+C_{fXR}^2\right) \left(m_X^2-m_f^2\right)+6 C_{fXL} C_{fXR} m_f^2 \right), \\
\Gamma(X \rightarrow \mathrm{hadron}) =& \Gamma(X \rightarrow \mu\bar{\mu})R_\mu^H(\sqrt{s}=m_X)  , \\
\mathrm{Br}(X \rightarrow f\bar{f}) =&\frac{\Gamma(X\rightarrow f\bar{f})}{\Gamma(X\rightarrow \mathrm{all})}, 
\end{align}
where $N_c$ is the color factor $3$ for quarks and $1$ for leptons. 
The coupling coefficients for each chirality of the final state fermion $C_{fXL}$ and $C_{fXR}$ are read from Appendix~\ref{sec:App:vertex}.
At the low mass range of $X$, 
 $\Gamma(X \rightarrow q\bar{q})$ is replaced with $\Gamma(X \rightarrow \mathrm{hadron}) $,
 where $R_\mu^H$ is the improved R-ratio given in Refs.~\cite{Ilten:2018crw,Foguel:2022ppx}.
The denominator $\Gamma(X\rightarrow \mathrm{all})$ is the total decay rate of the $X$ boson
 under the assumption that any scalar bosons and right-handed neutrinos are heavier than $X$.
We use the total decay rate of the $X$ boson including hadron final states of the dark photon and $Z'$ in the $U(1)_{B-L}$ 
model implemented in \texttt{FORESEE}~\cite{Kling:2021fwx} and given in Ref.\cite{Foguel:2022ppx}. 
The same $X \rightarrow \mathrm{hadron}$ decay width of the $U(1)_{B-L}$ model is also used as a substitute
 for that in the $U(1)_R$ model which is not available to our knowledge\footnote{
Though the neglected axial interaction might cause differences, which, however, should be at most by a factor. In addition, in a light mass region where $X$ decays into not quarks but hadrons, $X$ dominantly decays into leptons and experimental bounds are essentially determined by the travel distance as we will show. Thus, our calculation also provides a good approximation.
}. 
We show, as an example, the branching ratio of $X$ decay to charged lepton pair and neutrino pair
 for the minimal $U(1)_{B-L}$ in Figure~\ref{Fig:XdecayBr}.
%%%%%%%%%%%%%%%%%%%%
%%%%%%%%%%%%%%%%%%%

\subsection{Meson decay to the $X$ boson}

\begin{figure}[htbp]
    \begin{tabular}{ccc}
	\begin{minipage}{0.3\linewidth}
		\centering
		\includegraphics[width=0.9\textwidth]{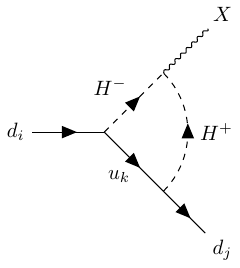}
		\vspace{-2mm}
		\subcaption*{(a)}
	\end{minipage} &
	\begin{minipage}{0.3\linewidth}
		\centering
		\includegraphics[width=0.9\textwidth]{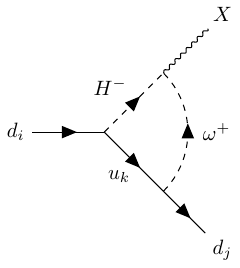}
		\vspace{-2mm}
		\subcaption*{(b)}
	\end{minipage} &
	\begin{minipage}{0.3\linewidth}
		\centering
		\includegraphics[width=0.9\textwidth]{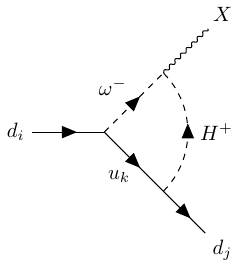}	
		\vspace{-2mm}
		\subcaption*{(c)}
	\end{minipage}   \\ 
	\begin{minipage}{0.3\linewidth}
		\centering
		\includegraphics[width=0.9\textwidth]{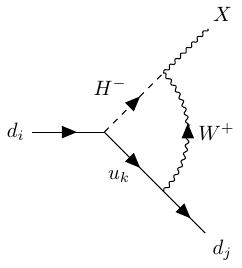}			
		\vspace{-2mm}
		\subcaption*{(d)}
	\end{minipage} &
	\begin{minipage}{0.3\linewidth}
		\centering
		\includegraphics[width=0.9\textwidth]{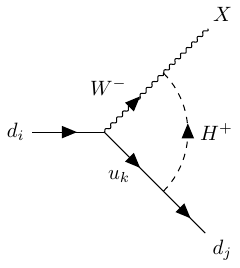}					
		\vspace{-2mm}
		\subcaption*{(e)}
	\end{minipage} &
	\begin{minipage}{0.3\linewidth}
		\centering
		\includegraphics[width=0.9\textwidth]{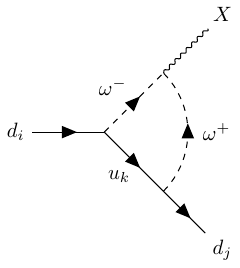}
		\vspace{-2mm}
		\subcaption*{(f)}
	\end{minipage}  \\ 
	\begin{minipage}{0.3\linewidth}
		\centering
		\includegraphics[width=0.9\textwidth]{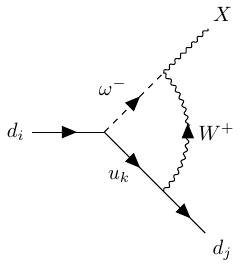}
		\vspace{-2mm}
		\subcaption*{(g)}
	\end{minipage} &
	\begin{minipage}{0.3\linewidth}
		\centering
		\includegraphics[width=0.9\textwidth]{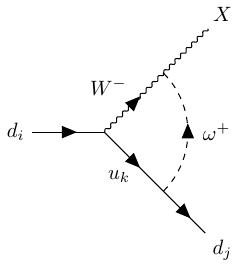}
		\vspace{-2mm}
		\subcaption*{(h)}
	\end{minipage} &
	\begin{minipage}{0.3\linewidth}
		\centering
		\includegraphics[width=0.9\textwidth]{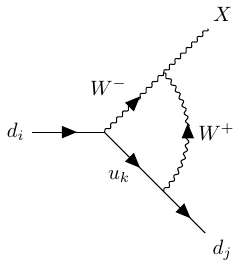}
		\vspace{-2mm}
		\subcaption*{(i)}
		\end{minipage}	 \\ 
	\begin{minipage}{0.3\linewidth}
		\centering
		\includegraphics[width=0.9\textwidth]{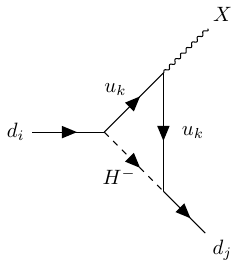}
		\vspace{-2mm}
		\subcaption*{(j)}
	\end{minipage} &
	\begin{minipage}{0.3\linewidth}
		\centering
		\includegraphics[width=0.9\textwidth]{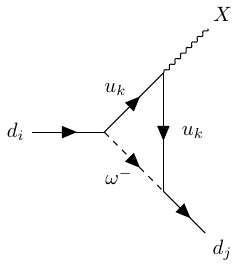}
		\vspace{-2mm}
		\subcaption*{(k)}
	\end{minipage} &
	\begin{minipage}{0.3\linewidth}
		\centering
		\includegraphics[width=0.9\textwidth]{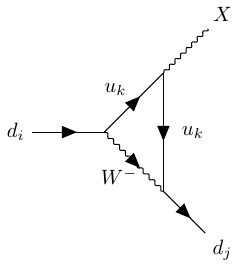}		
		\vspace{-2mm}
		\subcaption*{(l)}
		\end{minipage} \\
\end{tabular}
\caption{
Vertex correction diagrams induce $d_i \rightarrow d_j X$.
}
\label{Fig:diagram1}
\end{figure}

\begin{figure}[htbp]
\begin{tabular}{cc}
	\begin{minipage}{0.45\linewidth}
		\centering
		\includegraphics[width=0.85\textwidth]{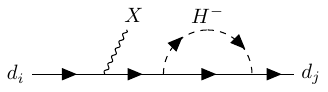}		
		\subcaption*{(m)}
	\end{minipage}  &
	\begin{minipage}{0.45\linewidth}
		\centering
		\includegraphics[width=0.85\textwidth]{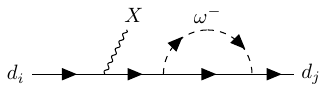}		
		\subcaption*{(n)}
	\end{minipage} \\ 
	\begin{minipage}{0.45\linewidth}
		\centering
		\includegraphics[width=0.85\textwidth]{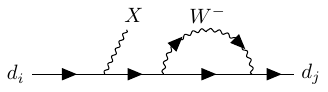}		
		\subcaption*{(o)}
	\end{minipage} &
	\begin{minipage}{0.45\linewidth}
		\centering
		\includegraphics[width=0.85\textwidth]{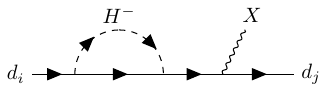}		
		\subcaption*{(p)}
	\end{minipage} \\ 
	\begin{minipage}{0.45\linewidth}
		\centering
		\includegraphics[width=0.85\textwidth]{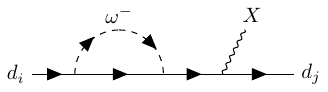}		
		\subcaption*{(q)}
	\end{minipage} &
	\begin{minipage}{0.45\linewidth}
		\centering
		\includegraphics[width=0.85\textwidth]{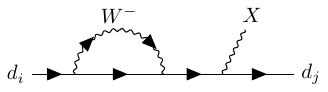}		
		\subcaption*{(r)}
	\end{minipage} \\
\end{tabular}
\caption{
Initial and final state radiation diagrams induce $d_i \rightarrow d_j X$.
}
\label{Fig:diagram2}
\end{figure}

Next, we consider two types of flavor changing meson decay into a lighter meson and
 the $X$ boson, $ B^- \rightarrow K^- X$ and $ K^- \rightarrow \pi^- X$. These meson decays occur 
 through flavor changing processes of down-type quarks $d_i$ into $d_j$ with the emission of $X$. 
The effective interaction Lagrangian of such flavor changing interactions $d_i \rightarrow d_j X$ is expressed as 
\begin{align}
\mathcal{L} \supset  \overline{d}_j \gamma^{\mu}\left(C_{d_jd_iXR}P_R +C_{d_jd_iXL} P_L\right)X_{\mu} d_i, \label{eq:lfv-coup}
\end{align}
which are induced by loop processes with the $W$ boson, the charged Higgs, NG bosons and up-type quarks\footnote{
Model-indenpendent expressions of the coupling  constants $C_{d_j d_i XR,L}$ are given in \cite{Altmannshofer:2023hkn}.}, 
as is shown in Figures~\ref{Fig:diagram1} and \ref{Fig:diagram2}. 
In general, loop-induced coupling constants are not finite and need the renormalization. 
The coefficients in Eq.~\eqref{eq:lfv-coup} become finite after putting together all amplitudes shown in Figure~\ref{Fig:diagram1} and \ref{Fig:diagram2}. 
There are two types of divergent terms; terms depend on the mass of an up-type quark $m_{u_k}$ propagating in the loop and the others do not.
The up-type quark mass dependent divergent terms cancel between them in the total amplitudes, while the mass-independent ones vanish for $d_i \neq d_j$ due to the unitarity relation $\sum_{u_k} V^{\dagger}_{ d_j  u_k}V_{u_k d_i} $ of the Cabbibo-Kobayashi-Maskawa (CKM) matrix elements $V_{u d}$.
The remaining finite terms have the form of $\sum_{u_k} V^\dagger _{d_j u_k }V_{u_k d_i} m_{u_k}^2/m_W^2$
 suppressed by not only usual loop factors but also the off-diagonal elements of the CKM matrix.  
For lighter quarks, the coefficients are further suppressed by the factor $m_{u_k}^2/m_W^2$. 
It would be worth noting that there are no terms solely proportional to $\log(m_{u_k}/m_W)$ for lighter quarks, which could give significantly enhanced contributions in some extended models~\cite{Marciano:1977cj,Raidal:1997hq,Gabrielli:2000te}, because the coefficients are induced by the $W^\pm$ and $H^\pm$ bosons.

In the following analyses, we employ the approximation of $m_{d_i (d_j)} =0$. We find that $C_{d_jd_iXR}$ vanishes under the approximation  
because only left-handed quarks interact with the charged Higgs boson as well as the weak boson. 
We present the detail of those derivation in Appendix~\ref{sec:AppC}. 
 
\subsubsection{$ B^- \rightarrow K^- X$}
\label{sec:BtoKX}

For the decay $B^-(q_B) \rightarrow K^-(q_K) X(q_X)$ where $q_i$ are the momentum of $i$-th species particle with the mass $m_i$,
the decay rate is given by  
\begin{align}
\Gamma(B^-\rightarrow K^-X) =& \frac{|\mathbf{q}_X|}{32\pi^2m_{B^-}^2}\int\left|\mathcal{M}(B^- \rightarrow K^- X)\right|^2d\Omega , \\
|\mathbf{q}_X|^2 = &\frac{m_{B^-}^4+\left(m_{K^-}^2-m_X^2\right)^2-2 m_{B^-}^2 \left(m_{K^-}^2+m_X^2\right)}{4 m_{B^-}^2} ,  \\
|\mathcal{M}(B^- \rightarrow K^- X)|^2 
 = & \frac{1}{4} f_K^2 \left|C_{sbXL}\right|^2 \left(\frac{\left(m_{B^-}^2-m_{K^-}^2\right)^2}{m_X^2}+m_X^2-2m_{B^-}^2-2m_{K^-}^2\right) . \label{eq:ampsq-B}
\end{align}
Here $f_K$ is the form factor of a kaon given as $\langle K^-|\bar{s}\gamma^{\mu}b|B^-\rangle \simeq f_K(q_B+q_K)^{\mu}$~\cite{Dreiner:2009er,Ball:2004ye}\footnote{A different form factor of the kaon has been reported based on lattice QCD results~\cite{Becirevic:2023aov}. By comparing those two, we find that the form factor shows $50$\% deviation near the threshold of $X$ prodcution. However, this uncertainty does not qualitatively affect our conclusion because this difference results in at most 1.4 times more stringent upper bounds of $\epsilon$ or $g_X$ near the threshold.}.
We note that the first term in Eq.~\eqref{eq:ampsq-B} originates from the longitudinal component of the $X$ boson and enhances the decay width in lighter $m_X$ region which plays an important role for the constraint from the rare meson decay.
The travel distance of $X$ produced by the $B$ meson decay with taking Lorentz $\gamma$ factor into account is estimated as
\begin{equation}
l_X = \sqrt{1+\frac{|\mathbf{q}_X|^2}{m_X^2}} \tau_X ,
\label{eq:lX}
\end{equation}
with $\tau_X=1/\Gamma(X \rightarrow \mathrm{all})$ being the lifetime of $X$ at the rest frame.

\subsubsection{$ K^- \rightarrow \pi^- X$}

As in the Sec.~\ref{sec:BtoKX}, we obtain the decay rate for $K^-(q_K) \rightarrow \pi^-(q_\pi) X(q_X)$ as
\begin{align} 
\Gamma(K^-\rightarrow \pi^-X) =& \frac{|\mathbf{q}_X|}{32\pi^2m_{K^-}^2}\int\left|\mathcal{M}(K^- \rightarrow \pi^- X)\right|^2d\Omega,  \\
|\mathbf{q}_X|^2 = & \frac{m_{K^-}^4+\left(m_{\pi^-}^2-m_X^2\right)^2-2 m_{K^-}^2 \left(m_{\pi^-}^2+m_X^2\right)}{4 m_{K^-}^2} ,  \\
|\mathcal{M}(K^- \rightarrow \pi^- X)|^2 
 = & \frac{1}{4} f_{\pi}^2 \left|C_{dsXL}\right|^2 \left(\frac{\left(m_{K^-}^2-m_{\pi^-}^2\right)^2}{m_X^2}+m_X^2-2m_{K^-}^2-2m_{\pi^-}^2\right) ,  \label{eq:ampsq-K}
\end{align}
where the form factor is quoted from Refs.~\cite{Dreiner:2009er,Nam:2007fx}

%%%%%%%%%%%%%%%%%%%%%%%
\section{Results}
\label{sec:Results}
%%%%%%%%%%%%%%%%%%%%%%%

We derive the constraints on the coupling constant and the mass $m_X$ from flavor changing meson decay modes
 by comparing those experimental results. 
For $B^+ \rightarrow K^+\ell^+\ell^-$, the SM prediction is obtained as
$\mathrm{Br}(B^+ \rightarrow K^+\ell^+\ell^-)|_{\mathrm{SM}} = (6.5\pm 0.5)\times 10^{-7}$~\cite{Parrott:2022zte}.
In fact, the SM prediction is somewhat larger than experimental measurements, 
\begin{align} 
& \mathrm{Br}(B^+ \rightarrow K^+\ell^+\ell^-)|_{\mathrm{exp(PDG)}} = (4.7\pm 0.5)\times 10^{-7} , \\
& \mathrm{Br}(B^+ \rightarrow K^+\ell^+\ell^-)|_{\mathrm{exp(Belle~II)}} = (5.99^{+0.45}_{-0.43})\times 10^{-7},
\end{align} 
where we quote the results in PDG~\cite{ParticleDataGroup:2024cfk}, which is the average
 of BaBaR~\cite{BaBar:2008jdv}, LHCb~\cite{LHCb:2012juf} and Belle~\cite{BELLE:2019xld}, and Belle II(2021)~\cite{BELLE:2019xld}.
Thus, apparently, there are no rooms for new physics to contribute to it constructively. 
In the following, as the constraint, we impose the condition that the $X$ decay contribution to this mode
 should be smaller than the uncertainty of the SM prediction, that is
 \begin{align}
 \mathrm{Br}(B^+ \rightarrow K^+ (X \to \ell^+\ell^-)) < 0.5\times 10^{-7}. 
\end{align}

The SM prediction for $K^+ \rightarrow\pi^+\nu\bar{\nu}$ is obtained as 
$\mathrm{Br}(K^+ \rightarrow\pi^+\nu\bar{\nu})|_{\mathrm{SM}} = (7.73 \pm 0.61)\times 10^{-11}$~\cite{Brod:2021hsj}.
The experimental result is found as 
$\mathrm{Br}(K^+ \rightarrow\pi^+\nu\bar{\nu})|_{\mathrm{exp(PDG)}} = (1.14^{+0.40}_{-0.33})\times 10^{-10}$~\cite{ParticleDataGroup:2024cfk} which is the average of E949~\cite{E949:2008btt} and NA62~\cite{NA62:2021zjw}.
We impose the constraint as the $X$ decay contribution to this mode should be 
\begin{align}
 \mathrm{Br}(K^+ \rightarrow \pi^+ (X \to \nu \bar{\nu})) < 0.83 \times 10^{-10}, 
\end{align}
which corresponds to $1\sigma$ deviation between the errors of the experiments and prediction.

%%%%%%%%%%%%%%%%%%%%%%%%%
\subsection{Dark photon}
%%%%%%%%%%%%%%%%%%%%%%%%%
For the minimal dark photon model, we take $x_H = x_\Phi = 0$ with the limit of $v_1 \to 0$ and $g_X \to 0$ in $C_{d_j d_i X R, L}$. 
Then, the charged Higgs boson contributions are absent because $\Phi_1$ is decoupled from quarks. As we explained in the previous 
section, the dominant contributions to the coefficients come from the top quark loops. 
The leading terms of the coefficient in the limit of large $x_t \equiv m_t^2/m_W^2$ is given by
\begin{align}
C_{b s X L} & = -\frac{g_2^3 c_W \sin\theta }{192 \pi^2} V^\dagger_{st} V_{tb}  \bigg( 
29 - \frac{7  x_t}{2} + (10+3 x_t) \log x_t + 32 \log \left( \frac{m_W^2}{m_Z^2} \right)
\bigg),
\end{align}
where the renormalization scale is taken to be $m_Z$, and the gauge mixing angle is approximated as Eq.~\eqref{eq:theta-approx}. 
The coefficient is insensitive to the $X$ boson mass. 
Therefore, the $X$ boson mass dependence in the rare meson decay branching ratio comes from the meson form factor and kinematical factor 
in Eq.~ \eqref{eq:ampsq-B} and  \eqref{eq:ampsq-K}.

%
%%%%%%%%%%%%%%%%%%%%
\begin{figure}[tbp]
\centering
\includegraphics[clip,width=13.0cm]{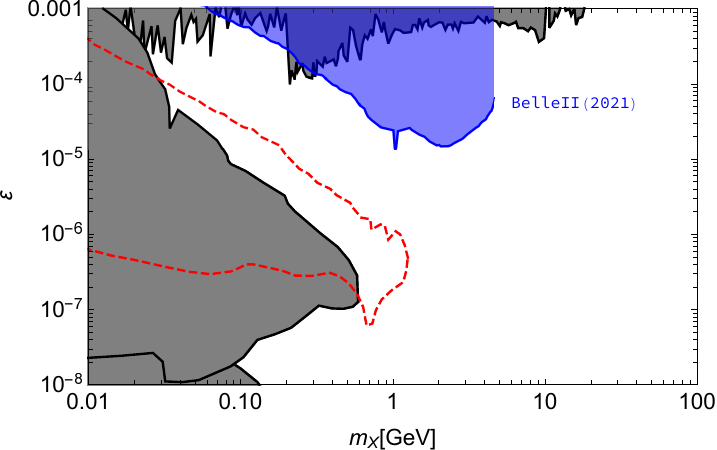}
\caption{
The constrained region by $B\rightarrow KX$ followed by $X\rightarrow \ell^+\ell^-$ is indicated by shading with blue. 
The other shaded regions are excluded by various beam dump, fixed target and collider experiments, and 
SN1987A. The future prospect for the FASER2 experiment is indicated by the red dashed curve.}
\label{Fig:DarkPhoton}
\end{figure}
%%%%%%%%%%%%%%%%%%%
%
The constraints for the minimal dark photon model are summarized in Figure \ref{Fig:DarkPhoton} of $m_X$ -- $\varepsilon$ plane. 
The bound from $B^+ \rightarrow K^+ (X \rightarrow \ell^+\ell^-) $ is indicated by bluish shaded region. 
The top gray shaded region is excluded mainly by the direct dark photon searches at BaBar and LHCb. 
In these searches, the dark photon is produced via $e^+ e^- \to \gamma X$ at BaBar \cite{BaBar:2014zli} and $q\bar{q} \to X$ at LHCb \cite{LHCb:2019vmc}, 
followed by the decays of $X \to e^+ e^-,~\mu^+ \mu^-$ inside the detector. The left gray region is excluded by beam dump 
experiments such as NuCal~\cite{Blumlein:2013cua}, E137~\cite{Marsicano:2018krp} and CHARM~\cite{Gninenko:2012eq}. 
These excluded regions are taken from Ref.~\cite{Bauer:2018onh} (related references therein).
The $B^+$ decay bound for $m_X \gtrsim 2$ GeV is given by the decay branching ratio Eq.~\eqref{eq:br-B},
while the bound for $m_X \lesssim 2$ GeV is determined by the travel distance, Eq.~\eqref{eq:prompt_decay} with Eq.~\eqref{eq:lX}. 
The decay branching ratio scales as $\varepsilon^2/m_X^2$ due to the longitudinal enhancement, and its constraint becomes stringent in lighter $m_X$ region. 
On the other hand,  the travel distance with the Lorentz factor is inversely proportional to $\varepsilon^2 m_X^2$, and favors 
larger $m_X$. These two constraints become competitive around $m_X = 2$ GeV, which determines the peak of the exclusion region.
We find that the parameter region around $m_X \sim 2$ GeV, $ 10^{-5} \lesssim \varepsilon \lesssim 10^{-3}$ is excluded by 
$B^+$ decay constraint. Note that there is no constraint from $K$ meson decay because dark photon has no interaction with neutrinos.
In addition, we add the bound obtained by FASER~\cite{FASER:2023tle}.
The future prospect for the FASER2 experiment is shown as the red dashed curve, taken from Ref.~\cite{Fabbrichesi:2020wbt}.

One can see that the constraints obtained from the rare meson decays exclude smaller $\varepsilon$ region than those from the direct searches (the top gray region). 
This results can be understood as follows: Firstly, the rare meson decay is a resonant two-body decay of $B \to K X$ followed by $X \to \ell^+ \ell^-$ while 
the SM process is three-body decay $B \to K \ell^+ \ell^-$. Thus, the $X$ contribution can be enhanced in terms of phase space.
Secondly, the branching ratio of $B \to K X$ is enhanced by $1/m_X^2$ due to the longitudinal component. These two enhancements result in stronger 
constraints than the direct searches.

%%%%%%%%%%%%%%%%%%%%%%%%%%%%%
\subsection{$U(1)_{B-L}$}
%%%%%%%%%%%%%%%%%%%%%%%%%%%%%
%
%%%%%%%%%%%%%%%%%%%%
\begin{figure}[tbp]
\centering
\includegraphics[clip,width=13.0cm]{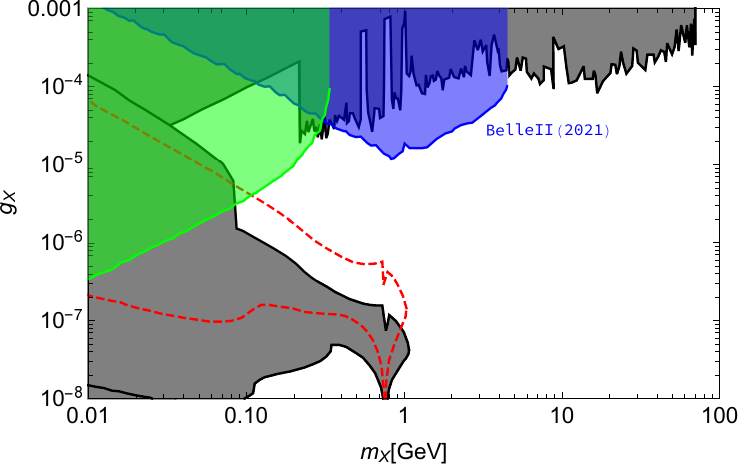}
\caption{
Constraints for the minimal $U(1)_{B-L}$ with $\epsilon =0$. 
The color code is the same as in Fig.~\ref{Fig:DarkPhoton}.}
\label{Fig:B-L}
\end{figure}
%%%%%%%%%%%%%%%%%%%

By taking $x_H = x_\Phi = 0$ and $v_1 \to 0$ with in $C_{d_j d_i X R, L}$, we obtain the minimal $U(1)_{B-L}$ model. The charged Higgs boson is absent as in the dark photon model. 
The leading term of $C_{b s X L}$ in the large $x_t$ is given by
\begin{align}
C_{b s X L} & = \frac{g_2^2 g_X}{192 \pi^2} V^\dagger_{st} V_{tb} 
\bigg(
7 + \frac{x_t}{2} - \log x_t + 4 \log \left( \frac{m_W^2}{m_Z^2} \right)
   \bigg).
\end{align}
Similarly to the dark photon model, the coefficient is insensitive to $m_X$.

The rare meson decay constraints for the minimal $U(1)_{B-L}$ model with $\epsilon =0$ are summarized
 in Figure \ref{Fig:B-L} with other experimental bounds.
The bound by $B^+ \rightarrow K^+ (X \rightarrow \ell^+\ell^-) $ is indicated by bluish shaded region,
while the greenish shaded region is excluded by $K^+ \rightarrow \pi^+ (X \rightarrow \nu\bar{\nu})$.
A similar explanation to the constraint from $B^+$ decay applies to the $K^+$ constraint. However, in this case, 
there is no requirement on the travel distance to $K^+$ for $X \to \nu\bar{\nu}$ decay. 
Thus, the exclusion region becomes more stringent in lighter $m_X$ region.
Those constraints exclude almost all parameter space of $g_X \gtrsim 10^{-5}$ and $m_X \lesssim 0.1$ GeV.
The decay mode $K^+ \rightarrow \pi^+ X$ imposes more stringent constraint than $B^+\rightarrow K^+X$ for
 $m_X \lesssim m_{K}$.
The constraints for the minimal $U(1)_{B-L}$ model is significant 
 for $g_X \gtrsim 10^{-5}$ for smaller $m_X$.
The other constraints by beam dump, fixed target, collider experiments and
 the future prospect for the FASER2 experiment are quoted from Refs.~\cite{Asai:2022zxw,Asai:2023mzl}.
Our plot also includes the bound obtained by FASER~\cite{FASER:2023tle}.

%%%%%%%%%%%%%%%%%%%%%%%%%%
\subsection{$U(1)_R$}
%%%%%%%%%%%%%%%%%%%%%%%%%%
%
%%%%%%%%%%%%%%%%%%%%
\begin{figure}[tbp]
\centering
\includegraphics[clip,width=13.0cm]{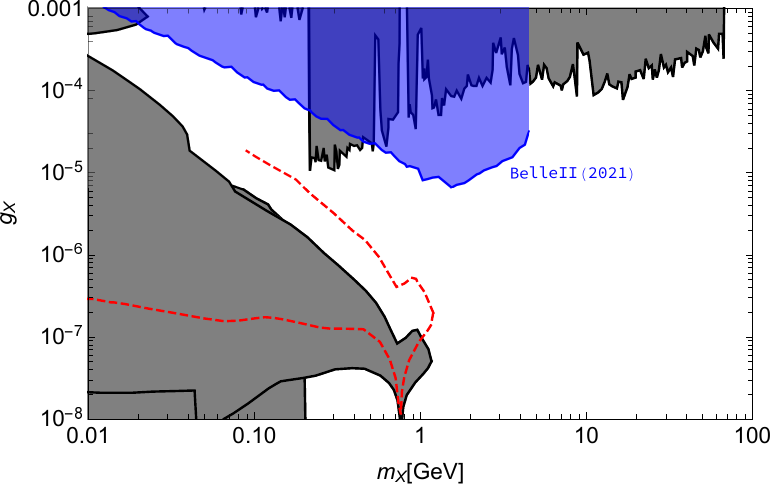}
\caption{
Constraints for the minimal $U(1)_R$ with $\epsilon =0$. 
The color code is the same as in Fig.~\ref{Fig:DarkPhoton}.
}
\label{Fig:R}
\end{figure}
%%%%%%%%%%%%%%%%%%%
%
This model can be reproduced by taking $x_H = -2$ and $x_\Phi = 2$. The leading terms of the coefficient in large $x_t$ and $x_{H^\pm} (= m_{H^\pm}^2/m_W^2)$ 
limit is given by
\begin{align}
C_{b s X L} & = \frac{g_2^2 g_X}{64 \pi^2} V^\dagger_{st} V_{tb} 
\bigg (
-4 + 5 x_t + (2 + 3 x_t) \log x_t -3 x_t \log x_{H^\pm}
\bigg).
\end{align}
The rare meson decay constraints for the minimal $U(1)_R$ model with $\epsilon =0$ are summarized
 in Fig.~\ref{Fig:R} with other experimental bounds.
As in the dark photon model, the constraints from the kaon decay is absent in the $U(1)_R$ model,
because the $U(1)_R$ gauge boson does not couple with (left-handed) neutrinos and the decay $X\rightarrow \nu\bar{\nu}$ does not occur. 
Thus, as a whole, the limits and its feature are similar to the case of dark photon.
The other constraints by beam dump, fixed target, collider experiments and
 the future prospect for the FASER2 experiment are quoted from Ref.~\cite{Asai:2022zxw}.
We show the constraint for $m_{H^+} = 500$ GeV in Fig.~\ref{Fig:R}.
We note that the result is almost insensitive for $m_{H^+}$, because
 the amplitudes is proportional to dimensionless $C_{24}$ of the Passarino Veltman function~\cite{Passarino:1978jh}\footnote{See App.~\ref{sec:Passarino Veltman functions} for our definition.}
 in processes mediated by the charged Higgs boson.
We also find that the longitudinal enhancement of the cross section in the $U(1)_R$ model reported in Ref.~\cite{Dror:2017nsg}
 does not exist in our models.

%%%%%%%%%%%%%%%%%%%%%%%
\section{Summary}
\label{sec:Summary}
%%%%%%%%%%%%%%%%%%%%%%%

We have evaluated the branching ratios of the flavor changing rare meson decays, $B^-\rightarrow K^- (X \rightarrow \ell^+\ell^-)$ and 
 $K^-\rightarrow \pi^- (X \rightarrow \nu\bar{\nu})$ through the on-shell vector boson $X$ production, and
 derived the corresponding bounds on some new physics models.
We explicitly show plots of the constraints for well-motivated examples of minimal dark photon, $U(1)_{B-L}$ and $U(1)_R$ models. 
In any case, it turns out for the gauge coupling constant and/or the gauge kinetic mixing angle to be smaller than about $10^{-5}$
 to be consistent with the rare meson decay.
Although the parameter range of $(m_X, g_X (\varepsilon)) \sim ( 0.1 \,\mathrm{GeV}, 10^{-5}-10^{-2}) $ has been regarded
 as unexplored and unconstrained, the $K^+ \rightarrow \pi^+ X$ bound actually fully excludes such parameter range
 in the $U(1)_{B-L}$ model.
The $B^+ \rightarrow K^+ X$ bound also imposes about one order of magnitude stringent bounds than by the other direct searches.

%======================================%
%<<<<<<<<<< ACKNOWLEDGMENTS >>>>>>>>>>>%
%======================================%

\section*{Acknowledgments}
This work is supported in part by KAKENHI Grants No.~JP23K03402 (O.S.), 22K03622 and 23K25885 (T.S.).

%======================================%
%<<<<<<<<<< APPENDIX >>>>>>>>>>>%
%======================================%
\appendix

%%%%%%%%%%%%%%%%%%%%%%%
\section{Higgs masses and spectrum}
\label{sec:App:Higgs}
%%%%%%%%%%%%%%%%%%%%%%%

For $x_{\Phi}=2$, stationary conditions are
\begin{subequations}
\begin{align}
& \frac{\lambda_{12} v_2 v_X}{\sqrt{2}}+\frac{\lambda_1 v_1^3}{2}+\frac{1}{2} \lambda_3 v_1 v_2^2+\frac{1}{2} \lambda_4 v_1 v_2^2+\hat{\mu_1}^2 v_1 =0, \\
& \frac{\lambda_{12} v_1 v_X}{\sqrt{2}}+\frac{\lambda_2 v_2^3}{2}+\frac{1}{2} \lambda_3 v_1^2 v_2+\frac{1}{2} \lambda_4 v_1^2 v_2-\hat{\mu_2}^2 v_2 =0, \\
& \frac{\lambda_{12} v_1 v_2}{\sqrt{2}}+\frac{\lambda_X v_X^3}{2}-\mu_X^2 v_X =0.
\end{align}
\end{subequations}

CP even Higgs boson mass matrix is
\begin{align}
 m_S^2=
\left(
\begin{array}{ccc}
 \lambda_1 v_1^2-\frac{\lambda_{12} v_2 v_X}{\sqrt{2} v_1} & \frac{\lambda_{12} v_X}{\sqrt{2}}+\lambda_3 v_1 v_2+\lambda_4 v_1 v_2 & \frac{\lambda_{12} v_2}{\sqrt{2}} \\
 \frac{\lambda_{12} v_X}{\sqrt{2}}+\lambda_3 v_1 v_2+\lambda_4 v_1 v_2 & \lambda_2 v_2^2-\frac{\lambda_{12} v_1 v_X}{\sqrt{2} v_2} & \frac{\lambda_{12} v_1}{\sqrt{2}} \\
 \frac{\lambda_{12} v_2}{\sqrt{2}} & \frac{\lambda_{12} v_1}{\sqrt{2}} & \lambda_X v_X^2-\frac{\lambda_{12} v_1 v_2}{\sqrt{2} v_X} \\
\end{array}
\right) .
\end{align}

Charged Higgs boson mass matrix
\begin{align}
 m_{H^{\pm}}^2=\left(
\begin{array}{cc}
 -\frac{v_2 \left(\sqrt{2} \lambda_{12} v_X+\lambda_4 v_1 v_2\right)}{2 v_1} & \frac{1}{2} \left(\sqrt{2} \lambda_{12} v_X+\lambda_4 v_1 v_2\right) \\
 \frac{1}{2} \left(\sqrt{2} \lambda_{12} v_X+\lambda_4 v_1 v_2\right) & -\frac{v_1 \left(\sqrt{2} \lambda_{12} v_X+\lambda_4 v_1 v_2\right)}{2 v_2} \\
\end{array}
\right) , 
\end{align}
is diagonalized as
\begin{align}
 m_{H^{\pm}}^2{}^\mathrm{diag} &= \mathrm{diag}
\left( 0,-\frac{v^2\left(\sqrt{2} \lambda_{12} v_X+\lambda_4 v_1 v_2\right)}{2 v_1 v_2}\right) = U_C^{\dagger} m_{H^{\pm}}^2 U_C ,
\end{align}
 with
\begin{align}
U_C = \left(
\begin{array}{cc}
 \frac{v_1}{v} & -\frac{v_2}{v} \\
 \frac{v_2}{v} & \frac{v_1}{v} \\
\end{array}
\right) . \label{eq:U_C}
\end{align}

CP odd Higgs boson mass matrix
\begin{align}
 m_P^2=
 \left(
\begin{array}{ccc}
 -\frac{\lambda_{12} v_2 v_X}{\sqrt{2} v_1} & \frac{\lambda_{12} v_X}{\sqrt{2}} & \frac{\lambda_{12} v_2}{\sqrt{2}} \\
 \frac{\lambda_{12} v_X}{\sqrt{2}} & -\frac{\lambda_{12} v_1 v_X}{\sqrt{2} v_2} & -\frac{\lambda_{12} v_1}{\sqrt{2}} \\
 \frac{\lambda_{12} v_2}{\sqrt{2}} & -\frac{\lambda_{12} v_1}{\sqrt{2}} & -\frac{\lambda_{12} v_1 v_2}{\sqrt{2} v_X} \\
\end{array}
\right)
\end{align}
diagonalized as
\begin{align}
 m_P^2{}^\mathrm{diag} &= \mathrm{diag}
\left( 0,0,-\frac{\lambda_{12} \left(v_1^2 v_2^2+v^2 v_X^2\right)}{\sqrt{2} v_1 v_2 v_X}\right) = U_P^{\dagger} m_P^2 U_P , \end{align}
 with
\begin{align}
U_P = 
\left(
\begin{array}{ccc}
 \frac{v_1}{v} & \frac{v_1 v_2^2}{v \sqrt{v_1^2 v_2^2+v^2 v_X^2}} & -\frac{v_2 v_X}{\sqrt{v_1^2 v_2^2+v^2 v_X^2}} \\
 \frac{v_2}{v} & -\frac{v_1^2 v_2}{v \sqrt{v_1^2 v_2^2+v^2 v_X^2}} & \frac{v_1 v_X}{\sqrt{v_1^2 v_2^2+v^2 v_X^2}} \\
 0 & \frac{v_X v}{ \sqrt{v_1^2 v_2^2+v^2 v_X^2}} & \frac{v_1 v_2}{\sqrt{v_1^2 v_2^2+v^2 v_X^2}} \\
\end{array}
\right) . \label{eq:U_P}
\end{align}

%%%%%%%%%%%%%%%%%%%%%%%
\section{Vertices}
\label{sec:App:vertex}
%%%%%%%%%%%%%%%%%%%%%%%

In the following, vertexes involving right-handed neutrinos $\nu_R$ are omitted.

\subsection{Fermion-fermion-scalar}
\label{subsec:Yukawa}

Yukawa interactions are
\begin{align}
 \mathcal{L}_Y =& - \overline{d_i}m_{d_i}\left(1+\frac{\phi_2}{v_2}\right)d_i- \overline{u_i}m_{u_i}\left(1+\frac{\phi_2}{v_2}\right)u_i- \overline{l_i}m_{l_i}\left(1+ \frac{\phi_2}{v_2}\right)l_i \nonumber \\
 & -i \frac{m_{d_i}}{v_2}\overline{d_i}a_2 \gamma_5 d_i - i \frac{m_{u_i}}{v_2}\overline{u_i}(-a_2)\gamma_5 u_i -i \frac{m_{l_i}}{v_2} \overline{l_i} a_2\gamma_5  l_i \nonumber\\
 & -\overline{u_i}V_{u_i d_j}\left( \frac{\sqrt{2}m_{d_j}}{v_2}P_R- \frac{\sqrt{2}m_{u_i}}{v_2} P_L\right)  \phi_2^+ d_j  -\frac{\sqrt{2}m_{l_i}}{v_2}\overline{\nu_i}\phi_2^+ P_R l_i   \nonumber \\ 
 & -\overline{d_i}V^{\dagger}_{d_i u_j} \left( \frac{\sqrt{2}m_{d_i}}{v_2}P_L- \frac{\sqrt{2}m_{u_j}}{v_2} P_R\right) \phi_2^- u_j -\frac{\sqrt{2}m_{l_i}}{v_2} \overline{l_i}\phi_2^-P_L \nu_i   ,
\end{align}
where $\phi_2, a_2$ and $\phi_2^{\pm}$ are related with physical states and NG modes through unitary matrices as in Eqs.~(\ref{eq:mixing:h}), (\ref{eq:mixing:H+}) and (\ref{eq:mixing:A}).

\subsection{Fermion-fermion-gauge}
\label{subsec:fermionGauge}

The gauge interactions for fermions are given by
\begin{align}
\mathcal{L}_{CC} =&  \frac{g_2}{\sqrt{2}}\overline{\nu_i}\gamma^{\mu}W^+_{\mu}P_L l_i 
 + \frac{g_2}{\sqrt{2}}\overline{l_i}\gamma^{\mu}W^-_{\mu}P_L \nu_i \nonumber \\
&  +\frac{g_2}{\sqrt{2}}\overline{u_i}\gamma^{\mu}V_{u_i d_j}W^+_{\mu}P_L d_j 
 + \frac{g_2}{\sqrt{2}}\overline{d_i}\gamma^{\mu}V^{\dagger}_{d_i u_j}W^-_{\mu}P_L u_j  ,
\end{align}
\begin{align}
\mathcal{L}_{EM} =& (-1)g_2s_W\overline{l_i}\gamma^{\mu}A_{\mu}l_i + \frac{2}{3}g_2s_W\overline{u_i}\gamma^{\mu}A_{\mu}u_i +\frac{-1}{3}g_2 s_W \overline{d_i}\gamma^{\mu}A_{\mu}d_i  ,
\end{align}
\begin{align}
\mathcal{L}_{NC}^l =& \frac{ g_2 c_{\epsilon} \left(c_{\theta} -s_{\theta} s_W t_{\epsilon}\right)- g_X c_W s_{\theta} (x_H+2)}{2 c_{\epsilon} c_W} \overline{\nu_i}\gamma^{\mu}Z_{\mu} P_L \nu_i  \nonumber \\
& +\frac{g_2 c_{\epsilon} \left(s_W (s_{\theta} t_{\epsilon}-c_{\theta} s_W)+c_{\theta} c_W^2\right)+g_X c_W s_{\theta} (x_H+2)}{2 c_{\epsilon} c_W}\overline{l_i}(-1)\gamma^{\mu}Z_{\mu}P_L l_i   \nonumber \\
& + \frac{g_2 c_{\epsilon} s_W (c_{\theta} s_W-s_{\theta} t_{\epsilon})-g_X c_W s_{\theta} (x_H+1)}{c_Wc_{\epsilon}}  \overline{l_i}\gamma^{\mu}Z_{\mu}P_R l_i ,
\end{align}
\begin{align}
\mathcal{L}_{NC}^q =& \frac{ g_2 c_{\epsilon} \left(s_W (s_{\theta} t_{\epsilon}-c_{\theta} s_W)+3 c_{\theta} c_W^2\right)+g_X c_W s_{\theta} (x_H+2)}{6 c_{\epsilon} c_W}\overline{u_i}\gamma^{\mu}Z_{\mu}P_L u_i   \nonumber \\
& +\frac{ g_2 2 c_{\epsilon} s_W (s_{\theta} t_{\epsilon}-c_{\theta} s_W)+g_X c_W s_{\theta} (2 x_H+1)}{3 c_{\epsilon} c_W}  \overline{u_i}\gamma^{\mu}Z_{\mu} P_R u_i             \nonumber \\
& +\frac{ g_2 c_{\epsilon} \left( s_W (s_{\theta}t_{\epsilon}-c_{\theta} s_W) -3c_{\theta}c_W^2\right)+g_X c_W s_{\theta} (x_H+2)}{6 c_{\epsilon} c_W}\overline{d_i}\gamma^{\mu}Z_{\mu}  P_L d_i    \nonumber \\
& +\frac{ g_2 (-1)c_{\epsilon} s_W (s_{\theta} t_{\epsilon}-c_{\theta} s_W)-g_X c_W s_{\theta} (x_H-1)}{3 c_{\epsilon} c_W}\overline{d_i}\gamma^{\mu}Z_{\mu}  P_R d_i     ,
\end{align}
\begin{align}
\mathcal{L}_{X}^l =& (-1)\frac{ g_2 c_{\epsilon} \left( c_{\theta} s_W t_{\epsilon}+s_{\theta} \right) +g_X c_{\theta} c_W (x_H+2) }{2 c_{\epsilon} c_W} \overline{\nu_i}\gamma^{\mu}X_{\mu}P_L \nu_i     \nonumber \\
& +(-1)\frac{ g_2 c_{\epsilon}\left( s_W (c_{\theta} t_{\epsilon}+s_{\theta} s_W)- c_W^2 s_{\theta} \right) +g_X c_{\theta} c_W (x_H+2) }{2 c_{\epsilon} c_W} \overline{l_i}\gamma^{\mu}X_{\mu} P_L l_i               \nonumber \\
& + (-1)\frac{ g_2 c_{\epsilon} s_W (c_{\theta} t_{\epsilon}+s_{\theta} s_W)+g_X c_{\theta} c_W (x_H+1)}{c_{\epsilon} c_W} \overline{l_i}\gamma^{\mu}X_{\mu}P_R l_i   ,
\end{align}
and
\begin{align}
\mathcal{L}_{X}^q =& \frac{ g_2 c_{\epsilon} \left( s_W (c_{\theta} t_{\epsilon}+s_{\theta} s_W)- 3c_W^2 s_{\theta}\right)+g_X c_{\theta} c_W (x_H+2) }{6 c_{\epsilon} c_W}\overline{u_i}\gamma^{\mu}X_{\mu}P_L u_i            \nonumber \\
& +\frac{g_2 c_{\epsilon}2s_W(c_{\theta}t_{\epsilon}+s_{\theta}s_W)+g_Xc_{\theta}c_W(2x_H+1)}{3c_{\epsilon}c_W}\overline{u_i}\gamma^{\mu}X_{\mu}P_R u_i              \nonumber \\
& +\frac{g_2 c_{\epsilon}\left( s_W(c_{\theta}t_{\epsilon}+s_{\theta}s_W)+ 3c_W^2s_{\theta}\right) +g_Xc_{\theta}c_W(x_H+2)}{6 c_{\epsilon} c_W} \overline{d_i}\gamma^{\mu}X_{\mu}P_L d_i            \nonumber \\
& +(-1)\frac{ g_2 c_{\epsilon} s_W (c_{\theta} t_{\epsilon}+s_{\theta} s_W)+g_X c_{\theta} c_W (x_H-1)}{3 c_{\epsilon} c_W} \overline{d_i}\gamma^{\mu}X_{\mu} P_R d_i  .
\end{align}

\subsection{Gauge-gauge-scalar}
\begin{align}
\mathcal{L}_{XW} =
& \frac{g_2}{2c_{\epsilon}c_Wv}\left( g_2 c_{\epsilon} s_W v^2 (c_{\theta} t_{\epsilon}+s_{\theta} s_W)+g_X c_{\theta}c_W \left(v_1^2 2x_{\Phi}+v^2x_H\right) \right)X^{\mu}W^-_{\mu}  \hat{\omega}^+     \nonumber \\
&+ \frac{g_2}{2c_{\epsilon}c_Wv}\left(g_2 c_{\epsilon}s_Wv^2 (c_{\theta} t_{\epsilon}+s_{\theta} s_W) +g_X c_{\theta} c_W \left(v_1^22x_{\Phi}+v^2x_H\right)\right)X^{\mu}W^+_{\mu}\hat{\omega}^-      \nonumber \\
& -\frac{g_2 g_X}{c_{\epsilon}v} c_{\theta}v_1 v_2 x_{\Phi} X^{\mu} \left( W^-_{\mu} H^+ + W^+_{\mu} H^- \right) .
\end{align}

\subsection{Gauge-scalar(NG)-scalar(NG)}
\begin{align}
\mathcal{L}_{X\phi\phi} =& -i \frac{g_X c_{\theta}v_1 v_2x_{\Phi}}{c_{\epsilon} v^2}  X^{\mu}(\partial_{\mu}H^+)\omega^-      \nonumber \\
 & -\frac{i}{2 c_{\epsilon} c_W } \left( g_2 c_{\epsilon} \left(c_W^2 s_{\theta}-s_W (c_{\theta} t_{\epsilon}+s_{\theta} s_W)\right)-g_X c_{\theta} c_W \left(x_H +\frac{2v_2^2}{v^2}x_{\Phi}\right) \right) X^{\mu}(\partial_{\mu}H^+) H^-      \nonumber \\
 & +i \frac{g_X c_{\theta}v_1 v_2x_{\Phi}}{c_{\epsilon} v^2}  X^{\mu}(\partial_{\mu}H^-)\omega^+      \nonumber \\
 & +\frac{i}{2 c_{\epsilon} c_W } \left( g_2 c_{\epsilon} \left(c_W^2 s_{\theta}-s_W (c_{\theta} t_{\epsilon}+s_{\theta} s_W)\right)-g_X c_{\theta} c_W \left(x_H +\frac{2v_2^2}{v^2}x_{\Phi}\right) \right) X^{\mu}(\partial_{\mu}H^-) H^+      \nonumber \\
& -i\frac{g_X c_{\theta} v_1 v_2x_{\Phi} }{c_{\epsilon} v^2} X^{\mu}(\partial_{\mu}\omega^+) H^-         \nonumber \\ 
& -\frac{i}{2 c_{\epsilon} c_W}\left( g_2 c_{\epsilon} \left(c_W^2 s_{\theta}-s_W (c_{\theta} t_{\epsilon}+s_{\theta} s_W)\right)-g_X c_{\theta} c_W \left(x_H +\frac{2v_1^2}{v^2}x_{\Phi}\right)  \right) X^{\mu}(\partial_{\mu}\omega^+) \omega^-             \nonumber \\ 
& +i\frac{g_X c_{\theta} v_1 v_2x_{\Phi} }{c_{\epsilon} v^2} X^{\mu}(\partial_{\mu}\omega^-) H^+         \nonumber \\ 
& +\frac{i}{2 c_{\epsilon} c_W}\left( g_2 c_{\epsilon} \left(c_W^2 s_{\theta}-s_W (c_{\theta} t_{\epsilon}+s_{\theta} s_W)\right)-g_X c_{\theta} c_W \left(x_H +\frac{2v_1^2}{v^2}x_{\Phi}\right)  \right) X^{\mu}(\partial_{\mu}\omega^-) \omega^+    .      
\end{align}

\subsection{Gauge-gauge-gauge}
\begin{align}
\mathcal{L}_{AWW} =& - g_2 s_W \left(\left(\partial_{\nu}W^-_{\mu}-\partial_{\mu}W^-_{\nu}
\right) W^+{}^{\nu}+ \left( \partial_{\mu}W^+_{\nu} -\partial_{\nu}W^+_{\mu} \right) W^-{}^{\nu}\right) A^{\mu}               \nonumber \\
& + g_2 s_W \left(W^-_{\mu} W^+_{\nu}-W^-_{\nu} W^+_{\mu}\right) \partial^{\nu}A^{\mu}    ,         
\end{align}
\begin{align}
\mathcal{L}_{ZWW} = & c_W g_2 c_{\theta}\left(\left(\partial_{\nu}W^-_{\mu}-\partial_{\mu}W^-_{\nu}\right) W^+{}^{\nu}+\left( \partial_{\mu}W^+_{\nu}-\partial_{\nu}W^+_{\mu} \right) W^-{}^{\nu}\right) Z^{\mu}             \nonumber \\
& + c_W g_2 c_{\theta}\left(W^-_{\nu} W^+_{\mu}-W^-_{\mu} W^+_{\nu}\right) \partial^{\nu}Z^{\mu} ,
\end{align}
\begin{align}
\mathcal{L}_{XWW} = & c_W g_2 s_{\theta}\left(\left(\partial_{\nu}W^-_{\mu}-\partial_{\mu}W^-_{\nu}\right) W^+{}^{\nu}+\left( \partial_{\mu}W^+_{\nu}-\partial_{\nu}W^+_{\mu} \right) W^-{}^{\nu}\right)  X^{\mu}             \nonumber \\
& + c_W g_2 s_{\theta}\left(W^-_{\nu} W^+_{\mu}-W^-_{\mu} W^+_{\nu}\right) \partial^{\nu}X^{\mu} .
\end{align}

\subsection{Gauge fixing}
\label{subsec:GaugeFixing}

We take the gauge fixing terms for Eq.~(\ref{Lag:gauge}) as
\begin{align}
 \mathcal{L}_\mathrm{g.f} &= -\frac{1}{2}G^2, \\
 G & =\frac{1}{\sqrt{\xi}}\left(\partial_{\mu}A^{a\mu}-\xi F^{a i} \omega_{i} \right) , 
\end{align}
 with the parameter $\xi$, 
 where $a$ runs $1,2,3,Y,X$, and 
\begin{subequations}
\begin{align}
\hat{\omega}^{\pm} &= \frac{v_1}{v}\phi_1^\pm+\frac{v_2}{v}\phi_2^{\pm}  ,  \\
\hat{\omega}_z &= \frac{v_1}{v}a_1+\frac{v_2}{v}a_2  ,  \\
\hat{\omega}_x &= \frac{v_1v_2^2 a_1 - v_1^2 v_2a_2+ v^2 v_X a_X}{v\sqrt{v^2 v_X^2+v_1^2 v_2^2}} ,
\end{align}
\end{subequations}
 are four NG bosons eaten by $W^{\pm}, Z$ and $X$. Note that those neutral $\hat{\omega}$ are not mass eigenstates of NG modes.
$F^{a i}$ are found, as so that the cross terms in
\begin{align}
& \mathcal{L}_2-\frac{1}{2}G^2  \nonumber \\
\supset & \frac{1}{2} ( v_1\partial_{\mu}a_1+ v_2\partial_{\mu}a_2) ( g_1 B^{\mu}-g_2 W^3{}^{\mu}) +\partial_{\nu}B^{\nu} F_{43} \hat{\omega}_z+\partial_{\nu}W^3{}^{\nu} F_{33} \hat{\omega}_z\nonumber \\
 &+ \left\{ v_1 \left( \frac{x_H}{2} + x_{\Phi}\right)\partial_{\mu}a_1+  v_2 \frac{x_H}{2}\partial_{\mu}a_2+2v_X \partial_{\mu}a_X \right\} g_X X^{\mu} +\partial_{\nu}X^{\nu} F_{54} \hat{\omega}_x \nonumber \\
 & +\partial_{\nu}B^{\nu} F_{44} \hat{\omega}_x+\partial_{\nu}W^3{}^{\nu} F_{34} \hat{\omega}_x+\partial_{\nu}X^{\nu} F_{53} \hat{\omega}_z ,
\end{align}
to be vanishing, for $x_{\Phi}=2$, as
\begin{subequations}
\begin{align}
F_{11}=& \frac{1}{2} (-g_2) v ,   \\
F_{22}=& \frac{1}{2} (-g_2) v  ,  \\
F_{33}=& \frac{1}{2} (-g_2) v ,   \\
F_{43}=& \frac{1}{2}g_1v ,  \\
F_{53}=& g_X \left(v \left(\frac{x_H}{2}+2 \right)-\frac{2 v_2^2}{v}\right),   \\
F_{54}=& 2 g_X \frac{\sqrt{v^2 v_X^2+v_1^2 v_2^2}}{v} ,  \\
\mathrm{others} =& 0 .
\end{align}
\end{subequations}
Mass squared for neutral NG bosons $\hat{\omega}_z$ and $ \hat{\omega}_x$,
\begin{align}
\left(
\begin{array}{cc}
 \frac{g_1^2 v^4+g_2^2 v^4+\left(g_X x_H v^2+4 g_X v_1^2\right)^2}{4 v^2} & \frac{g_X^2 \sqrt{v_1^2 v_2^2+v^2 v_X^2} \left(v_1^2 (x_H+4)+v_2^2 x_H\right)}{v^2} \\
 \frac{g_X^2 \sqrt{v_1^2 v_2^2+v^2 v_X^2} \left(v_1^2 (x_H+4)+v_2^2 x_H\right)}{v^2} & 4 g_X^2 \left(\frac{v_1^2 v_2^2}{v^2}+v_X^2\right) \\
\end{array}
\right) ,
\end{align}
is diagonalized to diag$(m_Z^2, m_X^2) $ for the mass eigenstates $\omega_z$ and $\omega_{z'}$.

%%%%%%%%%%%%%%%%%%%%%%%
\section{Evaluation of the $b \rightarrow s X$ vertices}
\label{sec:AppC}
%%%%%%%%%%%%%%%%%%%%%%%

We calculate the amplitude of $d_i(q_1) \rightarrow d_j(q_2) X(q_3)$,
 where $q_i$ are incoming and outgoing momentum, respectively.
This is induced by loop processes where up-type quarks, $W$, charged Higgs, and NG bosons running inside loops.
In the following calculations, terms proportional to mass of quarks, except for the top quark mass, are dropped.
Thus, dropped terms include those with couplings $C_{duH^-(\omega^-)L} \propto m_d$ and $C_{udH^+(\omega^+)R} \propto m_d$, and taking the form of $(\cdots)\slashed{q_1}$ and $\slashed{q_2}(\cdots)$ which can be rewritten to be so by using the Dirac equation $\slashed{q_1} u(q_1) \sim m_d u(q_1)$ and the conjugate of $\slashed{q_2} u(q_2) \sim m_d u(q_2)$. Some terms of $q_1^{\mu}\slashed{p}$ or $q_2^{\mu}\slashed{p}$ will be also such terms after integrated by the loop momentum $p$.
Terms proportional to $q_3^{\mu}$ are also omitted because those vanish
 when it is contracted with $-g^{\mu\nu}+q_3^{\mu}q_3^{\nu}/m_X^2$ due to the on-shell condition $q_3^2=m_X^2$.
Under this approximation, the transition amplitude takes the form of
\begin{align}
i\mathcal{M} = \epsilon^*_{\mu, X}(q_3)\overline{u}(q_2) \gamma^{\mu}i\left(C_{d_jd_iXR}P_R +C_{d_jd_iXL} P_L\right) u(q_1).
\end{align}
We have calculated all amplitudes in the $R_{\xi}$ gauge and
 confirmed that the total amplitude does not depends on the gauge parameter $\xi$.
Although we have taken the approximations mentioned above,
actually we have kept exceptionally some quark mass dependent terms in processes
of ISR, FSR and up-type quark emitting $X$ to check the gauge invariance and the cancellation of UV divergence.  
Nontrivial relations (\ref{eq:relations}) between coupling coefficients, as we will show, are used
 to see the cancellation of $\xi$ dependence.  
We note coefficients $i C_{d_jd_iXL}$ of all diagrams in Feynman-'t Hooft gauge ($\xi=1$) in Appendix~\ref{sec:amplitude}.

%Scalar(charge $q$, incoming momentum $p_1$)-scalar(outgoing momentum $p_2$)-gauge(gauge coupling $g$) vertex: $iqg(p_1+p_2)_{\mu}$

\subsection{Coupling List}

The abbreviated couplings used in Appendix~\ref{sec:amplitude} are defined as follows.
\begin{align}
&C_{d_j u_k H^- R}C_{u_k d_i H^+ L}C_{X(\partial H^-)H^+}  \nonumber \\
& =-\frac{2}{v^2}V^{\dagger}_{jk}m_{u_k}^2V_{ki}\left(\frac{v_1}{v_2}\right)^2\left(\frac{g_2}{2c_W}\left(s_Wc_{\theta}t_{\epsilon}-s_{\theta}(c_W^2-s_W^2)\right)+\frac{g_X c_{\theta}}{c_{\epsilon}}\left(\frac{1}{2}x_H +\frac{v_2^2}{v^2}x_{\Phi}\right)\right),   \\
&C_{X(\partial H^-)\omega^+}C_{d_j u_k \omega^- R} C_{u_k d_i H^+ L}= \frac{-2}{v^2}\left(\frac{v_1}{v_2}\right) V^{\dagger}_{jk}m_{u_k}^2V_{ki}\frac{-g_X c_{\theta} x_{\Phi}v_1 v_2 }{c_{\epsilon} v^2} ,   \\
&C_{X(\partial\omega^-)H^+}C_{d_j u_k H^- R}C_{u_k d_i \omega^+ L} = \frac{-2}{v^2}\left(\frac{v_1}{v_2}\right) V^{\dagger}_{jk}m_{u_k}^2V_{ki}\frac{-g_X c_{\theta} x_{\Phi}v_1 v_2 }{c_{\epsilon}v^2} ,   \\
&C_{d_j u_k W}m_{u_k}C_{u_k d_i H^+ L}C_{XWH^\pm} = \frac{g_2}{v}\left(\frac{v_1}{v_2}\right) V^{\dagger}_{d_j u_k} m_{u_k}^2V_{u_k d_i} \frac{-g_2v g_X c_{\theta} x_{\Phi} v_1 v_2}{c_{\epsilon}v^2}   ,   \\
&C_{d_j u_k H^- R}m_{u_k}C_{u_k d_i W}C_{XWH^\pm} =  \frac{g_2}{v}\left(\frac{v_1}{v_2}\right) V^{\dagger}_{d_j u_k} m_{u_k}^2V_{u_k d_i} \frac{-g_2v g_X c_{\theta} x_{\Phi} v_1 v_2}{c_{\epsilon}v^2}   ,   \\
&(-i)C_{X(\partial \omega^-)\omega^+}C_{d_j u_k \omega^- R}C_{u_k d_i \omega^+ L} \nonumber \\
& = V^{\dagger}_{jk}m_{u_k}^2V_{ki}\frac{-2}{v^2}\left( \frac{g_2}{2c_W}\left( s_Wc_{\theta}t_{\epsilon}-s_{\theta}(c_W^2-s_W^2) \right)+\frac{g_Xc_{\theta}}{c_{\epsilon}}\left(\frac{1}{2}x_H+\frac{v_1^2}{v^2}x_{\Phi}\right)  \right)  ,   \\
&C_{XW\omega^+}C_{d_j u_k W}m_{u_k}C_{u_k d_i \omega^+ L}   \nonumber \\
& = \frac{4}{v^2}V^{\dagger}_{d_j u_k} m_{u_k}^2V_{u_k d_i}
m_W^2\left(\frac{g_2}{2c_W}(s_Wc_{\theta}t_{\epsilon}+s_{\theta}s_W^2)+\frac{g_Xc_{\theta}}{c_{\epsilon}}\left(\frac{1}{2} x_H+x_{\Phi}\frac{v_1^2}{v^2}\right)\right)  ,   \\
&C_{XW\omega^-}C_{d_j u_k \omega^- R}m_{u_k}C_{u_k d_i W}   \nonumber \\
& = \frac{4}{v^2}V^{\dagger}_{d_j u_k} m_{u_k}^2V_{u_k d_i}
m_W^2\left(\frac{g_2}{2c_W}(s_Wc_{\theta}t_{\epsilon}+s_{\theta}s_W^2)+\frac{g_Xc_{\theta}}{c_{\epsilon}}\left(\frac{1}{2} x_H+x_{\Phi}\frac{v_1^2}{v^2}\right)\right) ,   \\
& C_{d_j u_k W}C_{u_k d_i W}C_{WWX} = \frac{g_2^2}{2}V^{\dagger}_{d_j u_k}V_{u_k d_i}g_2 c_W s_{\theta}  ,   \\
&C_{d_j u_k H^- R}C_{u_k d_i H^+ L}C_{u_k XR}  \nonumber \\
&=  \frac{2}{v^2}V^{\dagger}_{d_j u_k}m_{u_k}^2V_{u_jd_i }\left(\frac{v_1}{v_2}\right)^2 \frac{g_2 c_{\epsilon}2s_W(c_{\theta}t_{\epsilon}+s_{\theta}s_W)+g_Xc_{\theta}c_W(2x_H+1)}{3c_{\epsilon}c_W} ,   \\
&C_{d_j u_k H^- R}C_{u_k d_i H^+ L}C_{u_k XL}  \nonumber \\
& = \frac{2}{v^2}V^{\dagger}_{d_j u_k}m_{u_k}^2V_{u_kd_i } \left(\frac{v_1}{v_2}\right)^2 \frac{ g_2 c_{\epsilon} \left( s_W (c_{\theta} t_{\epsilon}+s_{\theta} s_W)- 3c_W^2 s_{\theta}\right)+g_X c_{\theta} c_W (x_H+2) }{6 c_{\epsilon} c_W} ,   \\
& C_{d_j u_k \omega^- R}C_{u_k d_i \omega^+ L}C_{u_k XR}
 = V^{\dagger}_{d_j u_k}m_{u_k}^2V_{u_jd_i }\frac{2}{v^2} \frac{g_2 c_{\epsilon}2s_W(c_{\theta}t_{\epsilon}+s_{\theta}s_W)+g_Xc_{\theta}c_W(2x_H+1)}{3c_{\epsilon}c_W} ,   \\
& C_{d_j u_k \omega^- R}C_{u_k d_i \omega^+ L}C_{u_k XL} 
 = V^{\dagger}_{d_j u_k}m_{u_k}^2V_{u_kd_i } \frac{2}{v^2} \frac{ g_2 c_{\epsilon} \left( s_W (c_{\theta} t_{\epsilon}+s_{\theta} s_W)- 3c_W^2 s_{\theta}\right)+g_X c_{\theta} c_W (x_H+2) }{6 c_{\epsilon} c_W} ,  % \\
\end{align}

\begin{align}
& C_{d_j u_k W}C_{u_k d_i W}C_{u_k XL} 
 = \frac{2 m_W^2}{v^2}V^{\dagger}_{d_j u_k}V_{u_k d_i}\frac{ g_2 c_{\epsilon} \left( s_W (c_{\theta} t_{\epsilon}+s_{\theta} s_W)- 3c_W^2 s_{\theta}\right)+g_X c_{\theta} c_W (x_H+2) }{6 c_{\epsilon} c_W}  ,  \\
& C_{d_j u_k W}C_{u_k d_i W}C_{u_k XR}  
 = \frac{2 m_W^2}{v^2}V^{\dagger}_{d_j u_k}V_{u_k d_i}\frac{g_2 c_{\epsilon}2s_W(c_{\theta}t_{\epsilon}+s_{\theta}s_W)+g_Xc_{\theta}c_W(2x_H+1)}{3c_{\epsilon}c_W} ,   \\
& C_{d_j u_k \omega^- R}C_{u_k d_i \omega^+ L}C_{u_k XR}=\frac{m_{u_k}^2}{m_W^2}C_{d_j u_k W}C_{u_k d_i W}C_{u_k XR} ,  \\
& C_{d_j u_k \omega^- R}C_{u_k d_i \omega^+ L}C_{u_k XL}=\frac{m_{u_k}^2}{m_W^2}C_{d_j u_k W}C_{u_k d_i W}C_{u_k XL} ,  \\
& C_{d_ju_k H^+R}C_{u_kd_iH^+L}C_{d_iXL}  \nonumber \\
& = V^{\dagger}_{d_j u_k}m_{u_k}^2V_{u_k d_i}\frac{2}{v^2} \left(\frac{v_1}{v_2}\right)^2\frac{g_2 c_{\epsilon}\left( s_W(c_{\theta}t_{\epsilon}+s_{\theta}s_W)+ 3c_W^2s_{\theta}\right) +g_Xc_{\theta}c_W(x_H+2)}{6 c_{\epsilon} c_W}  ,   \\
&C_{d_ju_k \omega^+R}C_{u_kd_i\omega^+L}C_{d_iXL} 
 = V^{\dagger}_{d_j u_k}m_{u_k}^2V_{u_k d_i}\frac{2}{v^2} \frac{g_2 c_{\epsilon}\left( s_W(c_{\theta}t_{\epsilon}+s_{\theta}s_W)+ 3c_W^2s_{\theta}\right) +g_Xc_{\theta}c_W(x_H+2)}{6 c_{\epsilon} c_W}  ,   \\
&C_{d_ju_k W}C_{u_kd_iW}C_{d_iXL}
=  \frac{g_2^2}{2}V^{\dagger}_{d_j u_k}V_{u_k d_i}\frac{g_2 c_{\epsilon}\left( s_W(c_{\theta}t_{\epsilon}+s_{\theta}s_W)+ 3c_W^2s_{\theta}\right) +g_Xc_{\theta}c_W(x_H+2)}{6 c_{\epsilon} c_W}   ,   \\ 
&C_{d_ju_k H^+R}C_{u_kd_iH^+L}C_{d_jXL} \nonumber \\
& = \frac{\partial \phi_2^-}{\partial H^-}\frac{\partial \phi_2^+}{\partial H^+}V^{\dagger}_{d_j u_k}\frac{2m_{u_k}^2}{v_2^2} V_{u_k d_i} \frac{g_2 c_{\epsilon}\left( s_W(c_{\theta}t_{\epsilon}+s_{\theta}s_W)+ 3c_W^2s_{\theta}\right) +g_Xc_{\theta}c_W(x_H+2)}{6 c_{\epsilon} c_W}   ,   \\
&C_{d_ju_k \omega^+R}C_{u_kd_i\omega^+L}C_{d_iXL} 
 = V^{\dagger}_{d_j u_k}m_{u_k}^2V_{u_k d_i}\frac{2}{v^2} \frac{g_2 c_{\epsilon}\left( s_W(c_{\theta}t_{\epsilon}+s_{\theta}s_W)+ 3c_W^2s_{\theta}\right) +g_Xc_{\theta}c_W(x_H+2)}{6 c_{\epsilon} c_W}  ,   \\
&C_{d_jXL}C_{d_ju_k W}C_{u_kd_iW} = \frac{g_2^2}{2} V^{\dagger}_{d_j u_k}V_{u_k d_i}  \frac{g_2 c_{\epsilon}\left( s_W(c_{\theta}t_{\epsilon}+s_{\theta}s_W)+ 3c_W^2s_{\theta}\right) +g_Xc_{\theta}c_W(x_H+2)}{6 c_{\epsilon} c_W} .
\end{align}

There are nontrivial relations between coupling coefficients as
\begin{subequations}
\begin{align}
& iC_{X(\partial \omega^-)\omega^+}C_{d_j u_k \omega^- R}C_{u_k d_i \omega^+ L} \left(\frac{2}{v^2}V^{\dagger}_{jk}m_{u_k}^2V_{ki}\right)^{-1}\nonumber \\
& = \frac{g_2}{2c_W}\left( s_Wc_{\theta}t_{\epsilon}-s_{\theta}(c_W^2-s_W^2) \right)+\frac{g_Xc_{\theta}}{c_{\epsilon}}\left(\frac{1}{2}x_H+\frac{v_1^2}{v^2}x_{\Phi}\right)    \nonumber\\
& = \left(1-\frac{m_X^2}{2m_W^2}\right)g_2 c_W s_{\theta} =\left(1-\frac{m_X^2}{2m_W^2}\right)C_{d_j u_k W}C_{u_k d_i W}C_{WWX} \left(\frac{g_2^2}{2}V^{\dagger}_{d_j u_k}V_{u_k d_i}\right)^{-1}  ,   \\
& C_{XW\omega^+}C_{d_j u_k W}m_{u_k}C_{u_k d_i \omega^+ L} \left(\frac{4}{v^2}V^{\dagger}_{d_j u_k} m_{u_k}^2V_{u_k d_i}
m_W^2\right)^{-1}  \nonumber \\
& = \frac{g_2}{2c_W}(s_Wc_{\theta}t_{\epsilon}+s_{\theta}s_W^2)+\frac{g_Xc_{\theta}}{c_{\epsilon}}\left(\frac{1}{2} x_H+x_{\Phi}\frac{v_1^2}{v^2}\right)    \nonumber \\
& = \left(1-\frac{m_X^2}{m_W^2}\right)\frac{1}{2}g_2 c_W s_{\theta} =\left(1-\frac{m_X^2}{m_W^2}\right)\frac{1}{2} C_{d_j u_k W}C_{u_k d_i W}C_{WWX} \left(\frac{g_2^2}{2}V^{\dagger}_{d_j u_k}V_{u_k d_i}\right)^{-1}.
\end{align}
\label{eq:relations}
\end{subequations}
In addition, relations
\begin{align}
C_{u_kd_i\omega^+ R}m_{u_k} = -C_{u_kd_i\omega^+ L}m_{d_i}, \\
C_{d_ju_k\omega^- L}m_{u_k} = -C_{d_ju_k\omega^- R}m_{d_j}, 
\end{align}
are taken into account, when we keep light quark masses.

\subsection{Coefficients in $\xi=1$ gauge}
\label{sec:amplitude}

Here, we note expression of $i C_{d_jd_iXL}$ of all diagrams whose name are same as
in Figs.~\ref{Fig:diagram1} and \ref{Fig:diagram2}.
The definition of Passarino Veltman functions are given in App.~\ref{sec:Passarino Veltman functions}.

\begin{enumerate}[label=(\alph*)\,]
%\subsubsection{$H^- u H^+$ $(p,q_1-p,p-q_3)$}
\item $H^- u H^+$ $(p,q_1-p,p-q_3)$
\begin{align}
& C_{d_j u_k H^- R}C_{u_k d_i H^+ L} C_{X(\partial H^-)H^+}
\int\frac{d^4 p}{(2\pi)^4}\frac{1}{p^2-m_{H^-}^2 }\frac{1}{(p-q_1)^2-m_{u_k}^2}\frac{1}{(p-q_3)^2-m_{H^-}^2 }2p^{\mu}\slashed{p} P_L \nonumber \\ 
= & \frac{i}{(4\pi)^2 Q^{4-D}}C_{d_j u_k H^- R}C_{u_k d_i H^+ L} C_{X(\partial H^-)H^+}2 C_{24}(q_1,q_3-q_1;m_{H^-}^2,m_{u_k}^2,m_{H^-}^2)\gamma^{\mu} P_L 
\end{align}
%

%\subsubsection{$H^-u\omega^+$}
\item $H^-u\omega^+$
\begin{align}
& C_{X(\partial H^-)\omega^+}C_{d_j u_k \omega^- R} C_{u_k d_i H^+ L}2\int\frac{d^4 p}{(2\pi)^4} \frac{1}{p^2-m_{H^-}^2 }\frac{1}{(p-q_1)^2-m_{u_k}^2} \frac{1}{(p-q_3)^2- m_W^2}\slashed{p}p^{\mu} P_L \nonumber \\
= & \frac{i}{(4\pi)^2 Q^{4-D}}C_{X(\partial H^-)\omega^+}C_{d_j u_k \omega^- R} C_{u_k d_i H^+ L}2C_{24}(q_1,q_3-q_1;m_{H^-}^2,m_{u_k}^2,m_W^2)\gamma^{\mu} P_L  ,
\end{align}

%\subsubsection{$\omega^- uH^+$}
\item $\omega^- uH^+$
\begin{align}
& C_{X(\partial \omega^-)H^+}C_{d_j u_k H^- R} C_{u_k d_i \omega^+ L} 2 \int\frac{d^4 p}{(2\pi)^4}\frac{1}{p^2-m_W^2}\frac{1}{(p-q_1)^2-m_{u_k}^2}\frac{1}{(p-q_3)^2-m_{H^+}^2}\slashed{p}p^{\mu} P_L \nonumber \\
=& \frac{i}{(4\pi)^2 Q^{4-D}}C_{X(\partial \omega^-)H^+}C_{d_j u_k H^- R} C_{u_k d_i \omega^+ L} 2C_{24}(q_1,q_3-q_1;m_W^2,m_{u_k}^2,m_{H^-}^2)\gamma^{\mu}  P_L  ,
\end{align}
%

%\subsubsection{$H^- uW^+$}
\item $H^- uW^+$
\begin{align}
& C_{d_j u_k W}m_{u_k}C_{u_k d_i H^+ L}C_{XWH^\pm}\int\frac{d^4 p}{(2\pi)^4}\frac{1}{p^2-m_{H^-}^2}\frac{1}{(p-q_1)^2-m_{u_k}^2}\frac{1}{(p-q_3)^2-m_W^2}\gamma^{\mu}  P_L  \nonumber \\
= & \frac{i}{(4\pi)^2 Q^{4-D}}C_{d_j u_k W}m_{u_k}C_{u_k d_i H^+ L}C_{XWH^\pm}C_{0}(q_1,q_3-q_1;m_{H^-}^2,m_{u_k}^2, m_W^2)\gamma^{\mu}  P_L ,
\end{align}
%

%\subsubsection{$W^- uH^+$}
\item $W^- uH^+$
\begin{align}
& C_{d_j u_k H^- R}m_{u_k}C_{u_k d_i W}C_{XWH^\pm}\int\frac{d^4 p}{(2\pi)^4}\frac{1}{p^2-m_W^2} \frac{1}{(p-q_1)^2-m_{u_k}^2} \frac{1}{(p-q_3)^2-m_{H^+}^2}\gamma^{\mu}P_L  \nonumber \\
= & \frac{i}{(4\pi)^2 Q^{4-D}}C_{d_j u_k H^- R}m_{u_k}C_{u_k d_i W}C_{XWH^\pm}C_{0}(q_1,q_3-q_1;m_W^2,m_{u_k}^2, m_{H^+}^2)\gamma^{\mu}P_L,
\end{align}
%

%\subsubsection{$\omega^-(q_1-p) u(p) \omega^+(q_2-p)$}
\item $\omega^-(q_1-p) u(p) \omega^+(q_2-p)$
\begin{align}
 & (-i)C_{X(\partial \omega^-)\omega^+}C_{d_j u_k \omega^- R}C_{u_k d_i \omega^+ L} \nonumber \\
 & \int\frac{d^4 p}{(2\pi)^4}\frac{1}{p^2-m_{u_k}^2}\frac{1}{(p-q_1)^2- m_W^2}\frac{1}{(p-q_2)^2-m_W^2}(2p-q_1-q_2)^{\mu}\slashed{p} P_L  \nonumber \\
= & \frac{i}{(4\pi)^2 Q^{4-D}}(-i)C_{X(\partial \omega^-)\omega^+}C_{d_j u_k \omega^- R}C_{u_k d_i \omega^+ L}2C_{24}(q_1,q_2-q_1;m_{u_k}^2,m_W^2,m_W^2)\gamma^{\mu}P_L  ,
\end{align}
%

%\subsubsection{$\omega^-(q_1-p) u(p)W^+(q_2-p)$}
\item $\omega^-(q_1-p) u(p)W^+(q_2-p)$
\begin{align}
& C_{XW\omega^+}C_{d_j u_k W}m_{u_k}C_{u_k d_i \omega^+ L} \int\frac{d^4 p}{(2\pi)^4}
\frac{1}{p^2-m_{u_k}^2}\frac{1}{(p-q_1)^2-m_W^2} \frac{1}{(p-q_2)^2-m_W^2}\gamma^{\mu} P_L  \nonumber \\
= & \frac{i}{(4\pi)^2 Q^{4-D}}C_{XW\omega^+}C_{d_j u_k W}m_{u_k}C_{u_k d_i \omega^+ L} C_0(q_1,q_2-q_1;m_{u_k}^2,m_W^2,m_W^2)\gamma^{\mu} P_L  , 
\end{align}
%

%\subsubsection{$W^-(q_1-p) u(p) \omega^+(q_2-p)$}
\item $W^-(q_1-p) u(p) \omega^+(q_2-p)$
\begin{align}
& C_{XW\omega^-}C_{d_j u_k \omega^- R}m_{u_k}C_{u_k d_i W} \int\frac{d^4 p}{(2\pi)^4} \frac{1}{p^2-m_{u_k}^2} \frac{1}{(p-q_1)^2-m_W^2}\frac{1}{(p-q_2)^2-  m_W^2}\gamma^{\mu} P_L  \nonumber \\
= & \frac{i}{(4\pi)^2 Q^{4-D}}C_{XW\omega^-}C_{d_j u_k \omega^- R}m_{u_k}C_{u_k d_i W} C_0(q_1,q_2-q_1;m_{u_k}^2,m_W^2,m_W^2)\gamma^{\mu} P_L   , 
\end{align}
%

%\subsubsection{$u(p)W^+_{\lambda\rho}(q_1-p)W^-_{\sigma\nu}(q_2-p) $}
\item $u(p)W^+_{\lambda\rho}(q_1-p)W^-_{\sigma\nu}(q_2-p) $
\begin{align}
& \int\frac{d^4 p}{(2\pi)^4}iC_{d_j u_k W}\gamma^{\sigma}P_L
\frac{i(\slashed{p}+m_{u_k})}{p^2-m_{u_k}^2} i C_{u_k d_i W}\gamma^{\rho} P_L\frac{-i}{(q_2-p)^2-m_W^2}\delta^{\nu}_{\sigma}  \nonumber \\
& iC_{WWX}\left( (q_2-q_3-p)_{\lambda}g_{\mu\nu}+(2p-q_1-q_2)_{\mu}g_{\nu\lambda}+(q_1+q_3-p)_{\nu}g_{\lambda\mu} \right) \frac{-i}{(q_1-p)^2-m_W^2}\delta^{\rho}_{\lambda} \nonumber \\ 
= & -\frac{i}{(4\pi)^2 Q^{4-D}} C_{d_j u_k W}C_{u_k d_i W}C_{WWX}2 \nonumber \\
& \left(  (m_{u_k}^2-2m_W^2) C_0(q_1,q_2-q_1;m_{u_k}^2,m_W^2,m_W^2)-2C_{24}(q_1,q_2-q_1;m_{u_k}^2,m_W^2,m_W^2) \right) \gamma^{\mu}P_L  \nonumber \\ 
& - \frac{i}{(4\pi)^2 Q^{4-D}}C_{d_j u_k W}C_{u_k d_i W}C_{WWX}2
\nonumber \\
& \left(B_0(q_2-q_1;m_W^2,m_W^2)-B_0(q_2;m_{u_k}^2,m_W^2)-B_0(q_1;m_{u_k}^2,m_W^2)\right)\gamma^{\mu} P_L  , %\nonumber \\ 
\end{align}
%

%\subsubsection{$u H^-u (p, q_1-p, p-q_1+q_2 )$}
\item $u H^-u (p, q_1-p, p-q_1+q_2 )$
\begin{align}
& C_{d_j u_k H^- R}C_{u_k d_i H^+ L}C_{u_k XR}\int\frac{d^4 p}{(2\pi)^4}\left(\frac{1}{p^2-m_{u_k}^2}\frac{1}{(p-q_1+q_2)^2-m_{u_k}^2}\gamma^{\mu}   \right.\nonumber \\
& \left. +\frac{1}{p^2-m_{H^+}^2 }\frac{1}{(p-q_1)^2-m_{u_k}^2}\frac{1}{(p-q_2)^2-m_{u_k}^2}(m_{H^+}^2\gamma^{\mu} -2 p^{\mu}\slashed{p} )  \right) P_L  \nonumber \\  
& -C_{d_j u_k H^- R}C_{u_k d_i H^+ L}C_{u_k XL}\int\frac{d^4 p}{(2\pi)^4} \frac{1}{p^2-m_{u_k}^2}\frac{1}{(q_1-p)^2-m_{H^+}^2 } \frac{1}{(p-q_1+q_2)^2-m_{u_k}^2} m_{u_k}^2\gamma^{\mu} P_L \nonumber \\
 = & \frac{i}{(4\pi)^2 Q^{4-D}}C_{d_j u_k H^- R}C_{u_k d_i H^+ L}C_{u_k XR}\left( B_0(q_1-q_2;m_{u_k}^2,m_{u_k}^2)    \right.\nonumber \\
& \left. +  m_{H^+}^2 C_0(q_1,q_2-q_1;m_{H^+}^2,m_{u_k}^2,m_{u_k}^2)  -2 C_{24}(q_1,q_2-q_1;m_{H^+}^2,m_{u_k}^2,m_{u_k}^2)  \right) \gamma^{\mu} P_L  \nonumber \\  
& -\frac{i}{(4\pi)^2 Q^{4-D}}C_{d_j u_k H^- R}C_{u_k d_i H^+ L}C_{u_k XL}C_0(q_1,-q_2;m_{u_k}^2,m_{H^+}^2,m_{u_k}^2)m_{u_k}^2\gamma^{\mu} P_L , 
\end{align}
%

%\subsubsection{$u \omega^- u$}
\item $u \omega^- u$
\begin{align}
& C_{d_j u_k \omega^- R}C_{u_k d_i \omega^+ L}C_{u_k XR}\int\frac{d^4 p}{(2\pi)^4}\left(\frac{1}{p^2-m_{u_k}^2}\frac{1}{(p-q_1)^2-m_W^2 }\frac{1}{(p-q_1+q_2)^2-m_{u_k}^2}(p^2-2p\!\cdot\!q_1)\gamma^{\mu}   \right.\nonumber \\
& \left. -2\frac{1}{p^2-m_{u_k}^2}\frac{1}{(p-q_1)^2-m_W^2 }\frac{1}{(p-q_1+q_2)^2-m_{u_k}^2}p^{\mu}\slashed{p}  \right) P_L  \nonumber \\  
 & -C_{d_j u_k \omega^- R}C_{u_k d_i \omega^+ L} C_{u_k XL}\int\frac{d^4 p}{(2\pi)^4} \frac{1}{p^2-m_{u_k}^2}\frac{1}{(p-q_1)^2-m_W^2 }\frac{1}{(p-q_1+q_2)^2-m_{u_k}^2}m_{u_k}^2\gamma^{\mu} P_L   \nonumber \\
 = & \frac{i}{(4\pi)^2 Q^{4-D}}C_{d_j u_k \omega^- R}C_{u_k d_i \omega^+ L}C_{u_k XR}    \nonumber \\ 
&
\left( B_0(q_1-q_2,m_{u_k}^2,m_{u_k}^2) + m_W^2 C_0(q_1,-q_2;m_{u_k}^2, m_W^2,m_{u_k}^2) -2C_{24}(q_1,-q_2;m_{u_k}^2, m_W^2,m_{u_k}^2) \right)  \gamma^{\mu} P_L  \nonumber \\  
 & -\frac{i}{(4\pi)^2 Q^{4-D}}C_{d_j u_k \omega^- R}C_{u_k d_i \omega^+ L} C_{u_k XL} C_0(q_1,-q_2;m_{u_k}^2, m_W^2,m_{u_k}^2) m_{u_k}^2\gamma^{\mu} P_L  ,
\end{align}
%

%\subsubsection{$u(p) W^-(q_1-p) u(p-q_1+q_2)$}
\item $u(p) W^-(q_1-p) u(p-q_1+q_2)$ 
\begin{align} 
& C_{d_j u_k W}C_{u_k d_i W}C_{u_k XR}(-2)m_{u_k}^2\int\frac{d^4 p}{(2\pi)^4}\frac{1}{p^2-m_{u_k}^2} \frac{1}{(p-q_1)^2-m_W^2}\frac{1}{(p-q_1+q_2)^2-m_{u_k}^2}\gamma^{\mu}P_L \nonumber \\
 & + C_{d_j u_k W}C_{u_k d_i W}C_{u_k XL}2\int\frac{d^4 p}{(2\pi)^4}\frac{1}{p^2-m_{u_k}^2} \frac{1}{(p-q_1)^2-m_W^2}\frac{1}{(p-q_1+q_2)^2-m_{u_k}^2} \nonumber \\
 &\left(\left( 2m_{u_k}^2-m_X^2-m_W^2\right)  \gamma^{\mu}-2p^{\mu} \slashed{p} \right) P_L     \nonumber \\
 & + C_{d_j u_k W}C_{u_k d_i W}C_{u_k XL}2\int\frac{d^4 p}{(2\pi)^4}    \nonumber \\
 & \left( \frac{1}{p^2-m_{u_k}^2}\left(  \frac{1}{(p-q_1)^2-m_W^2}-\frac{1}{(p-q_1+q_2)^2-m_{u_k}^2}\right) +\frac{1}{p^2-m_{u_k}^2} \frac{1}{(p-q_2)^2-m_W^2}  \right) \gamma^{\mu}  P_L     \nonumber \\
 = &  \frac{i}{(4\pi)^2 Q^{4-D}}C_{d_j u_k W}C_{u_k d_i W}C_{u_k XR}(-2)m_{u_k}^2 C_0(q_1,-q_2;m_{u_k}^2,m_W^2,m_{u_k}^2)\gamma^{\mu}P_L \nonumber \\
 & + \frac{i}{(4\pi)^2 Q^{4-D}}C_{d_j u_k W}C_{u_k d_i W}C_{u_k XL}2 \nonumber \\
 &\left(C_0(q_1,-q_2;m_{u_k}^2, m_W^2,m_{u_k}^2) \left( 2m_{u_k}^2-m_X^2-m_W^2\right) -2C_{24}(q_1,-q_2;m_{u_k}^2, m_W^2,m_{u_k}^2) \right) \gamma^{\mu} P_L     \nonumber \\
 & + \frac{i}{(4\pi)^2 Q^{4-D}}C_{d_j u_k W}C_{u_k d_i W}C_{u_k XL}2 \nonumber \\
 & \left(B_0(q_1,m_{u_k}^2,m_W^2)  -B_0(q_3,m_{u_k}^2,m_{u_k}^2) + B_0(q_2,m_{u_k}^2,m_W^2) \right) \gamma^{\mu}  P_L ,
\end{align}
%

%\subsubsection{ISR with $H^+$ loop}
\item ISR with $H^+$ loop
\begin{align}
& - C_{d_ju_k H^+R}C_{u_kd_iH^+L}C_{d_iXL}\frac{1}{(q_1-q_3)^2-m_{d_i}^2}2\int\frac{d^4 p}{(2\pi)^4} \frac{1}{p^2-m_{u_k}^2}\frac{1}{(q_2-p)^2-m_{H^+}^2}p\!\cdot\!q_2\gamma^{\mu}P_L     \nonumber \\
= & - \frac{i}{(4\pi)^2 Q^{4-D}}C_{d_ju_k H^+R}C_{u_kd_iH^+L}C_{d_iXL}\frac{1}{(q_1-q_3)^2-m_{d_i}^2} \nonumber \\
&\quad \times \big( q_2^2 B_1(q_2;m_{u_k}^2,m_{H^+}^2) - (q_1^2 + q_2^2) B_0(q_2;m_{u_k}^2,m_{H^+}^2) \big) \gamma^{\mu}P_L ,
\end{align}
%

%\subsubsection{ISR with $\omega^+$ loop}
\item ISR with $\omega^+$ loop 
\begin{align}
& - C_{d_ju_k \omega^+R}C_{u_kd_i\omega^+L}C_{d_iXL}\frac{1}{(q_1-q_3)^2-m_{d_i}^2}\int\frac{d^4 p}{(2\pi)^4} \frac{1}{p^2-m_{u_k}^2}\frac{ 1}{(p-q_2)^2-m_W^2}2p\!\cdot\!q_2 \gamma^{\mu}P_L   \nonumber \\
= & - \frac{i}{(4\pi)^2 Q^{4-D}}C_{d_ju_k \omega^+R}C_{u_kd_i\omega^+L}C_{d_iXL}\frac{1}{(q_1-q_3)^2-m_{d_i}^2} \nonumber \\
&\quad \times \big( q_2^2 B_1(q_2;m_{u_k}^2,m_W^2) - (q_1^2 + q_2^2) B_0(q_2;m_{u_k}^2,m_W^2) \big) \gamma^{\mu}P_L  ,
\end{align}
%

%\subsubsection{ISR with $W^+$ loop}
\item ISR with $W^+$ loop
\begin{align}
& C_{d_ju_k W}C_{u_kd_iW}C_{d_iXL}\frac{1}{(q_1-q_3)^2-m_{d_i}^2} \int\frac{d^4 p}{(2\pi)^4}\frac{1}{p^2-m_{u_k}^2} \frac{1}{(p-q_2)^2-m_W^2}(-2)2p\!\cdot\!q_2\gamma^{\mu}P_L    \nonumber  \\
= & \frac{i}{(4\pi)^2 Q^{4-D}}C_{d_ju_k W}C_{u_kd_iW}C_{d_iXL}\frac{1}{(q_1-q_3)^2-m_{d_i}^2} (-2) B_1(q_2;m_{u_k}^2,m_W^2)q_2^2\gamma^{\mu}P_L   ,
\end{align}
%

%\subsubsection{FSR with $H^+$ loop}
\item FSR with $H^+$ loop 
\begin{align}
& -C_{d_ju_k H^+R}C_{u_kd_iH^+L}C_{d_jXL} \gamma^{\mu}\frac{1}{q_1^2-m_{d_j}^2}
\int\frac{d^4 p}{(2\pi)^4}\frac{1}{p^2-m_{u_k}^2}\frac{1}{(p-q_1)^2-m_{H^+}^2}2p\!\cdot\!q_1 P_L     \nonumber \\
= & -\frac{i}{(4\pi)^2 Q^{4-D}}C_{d_ju_k H^+R}C_{u_kd_iH^+L}C_{d_jXL} \frac{1}{q_1^2-m_{d_j}^2}  \nonumber \\
&\quad \times \big( q_1^2 B_1(q_1;m_{u_k}^2,m_{H^+}^2) - (q_1^2 + q_2^2) B_0(q_1;m_{u_k}^2,m_{H^+}^2) \big) \gamma^{\mu}P_L    ,
\end{align}
%

%\subsubsection{FSR with $\omega^+$ loop}
\item FSR with $\omega^+$ loop 
\begin{align}
 & -C_{d_ju_k \omega^+R}C_{u_kd_i\omega^+L}C_{d_jXL} \gamma^{\mu}\frac{1}{q_1^2-m_{d_j}^2}
\int\frac{d^4 p}{(2\pi)^4}\frac{1}{p^2-m_{u_k}^2}\frac{1}{(p-q_1)^2- m_W^2}2p\!\cdot\!q_1 P_L   \nonumber \\
= &  -\frac{i}{(4\pi)^2 Q^{4-D}}C_{d_ju_k \omega^+R}C_{u_kd_i\omega^+L}C_{d_jXL} \frac{1}{q_1^2-m_{d_j}^2} \nonumber \\
&\quad \times \big(q_1^2 B_1(q_1;m_{u_k}^2,m_W^2) - (q_1^2 + q_2^2) B_0(q_1;m_{u_k}^2,m_W^2) \big) \gamma^{\mu}P_L   ,
\end{align}
%

%\subsubsection{FSR with $W^+$ loop}
\item FSR with $W^+$ loop
\begin{align}
& C_{d_jXL}C_{d_ju_k W}C_{u_kd_iW}(-4) \frac{1}{q_1^2-m_{d_j}^2}
\int\frac{d^4 p}{(2\pi)^4}\frac{1}{p^2-m_{u_k}^2}\frac{1}{(p-q_1)^2-m_W^2}p\!\cdot\!q_1\gamma^{\mu}P_L 
\nonumber \\
= & \frac{i}{(4\pi)^2 Q^{4-D}}C_{d_jXL}C_{d_ju_k W}C_{u_kd_iW} (-2) \frac{1}{q_1^2-m_{d_j}^2}B_1(q_1;m_{u_k}^2,m_W^2)q_1^2\gamma^{\mu}P_L   ,
\end{align}
\end{enumerate}

\section{Passarino Veltman functions}
\label{sec:Passarino Veltman functions}

We adopt the notation of Passarino Veltman function in Ref.~\cite{Pierce:1996zz} with Lorentzian signature\footnote{The original~\cite{Passarino:1978jh} is of Euclidian signature.} as 
\begin{align}
& \left[B_0, B^{\mu}, B^{\mu\nu} \right](p_1;m_1^2,m_2^2) %\nonumber \\
%& 
:= \frac{(4\pi)^2}{i}Q^{4-D}\int\frac{d^D k}{(2\pi)^D}\frac{\left[1,k^{\mu},k^{\mu}k^{\nu} \right]}{\left(k^2-m_1^2+i\epsilon\right)\left((k-p_1)^2-m_2^2+i\epsilon\right)} , \\
& \left[C_0, C^{\mu}, C^{\mu\nu}\right](p_1,p_2;m_1^2,m_2^2,m_3^2) \nonumber \\
& := \frac{(4\pi)^2}{i}Q^{4-D}\int\frac{d^D k}{(2\pi)^D}\frac{\left[1,k^{\mu},k^{\mu}k^{\nu}\right]}{\left(k^2-m_1^2+i\epsilon\right)\left((k-p_1)^2-m_2^2+i\epsilon\right)\left((k-p_1-p_2)^2-m_3^2+i\epsilon\right)} ,
\end{align}
with $D=4-2\epsilon$ and $Q$ being the renormalization scale.
$C$ functions can be decomposed as
\begin{align}
C^{\mu}&= p_1^{\mu} C_1+p_2^{\mu} C_2 ,\\
C^{\mu\nu}&= p_1^{\mu}p_1^{\nu}C_{21}+p_2^{\mu}p_2^{\nu}C_{22}+\{p_1p_2\}^{\mu\nu}C_{23}+g^{\mu\nu}C_{24} ,
\end{align}
and some useful analytic formulae are 
\begin{align}
A_0(m^2;Q)&=m^2\left( C_\mathrm{UV}+1-\log\left(\frac{m^2}{Q^2}\right)\right) ,\\
B^{\mu}(p;m_1^2,m_2^2)&= p^{\mu} B_1(p;m_1^2,m_2^2) ,\\
B_0(0;m_1^2,m_2^2;Q) &= 
\left\{
\begin{array}{ll}
C_\mathrm{UV}+1-\log\left(\frac{m_1^2}{Q^2}\right) +\frac{m_2^2}{m_1^2-m_2^2}\log\left(\frac{m_2^2}{m_1^2}\right)  & m_1 \neq m_2 \\
C_\mathrm{UV}-\log\left(\frac{m_1^2}{Q^2}\right) & m_1 = m_2
\end{array}
\right. , \\
2p^2 B_1(p;m_1^2,m_2^2;Q) &= A_0(m_2^2;Q)-A_0(m_1^2;Q)+(p^2+m_1^2-m_2^2)B_0(p;m_1^2,m_2^2;Q) ,\\
C_0(0,0;m_1^2,m_2^2,m_3^2) &= 
\left\{
\begin{array}{ll}
\frac{1}{m_2^2-m_3^2}\left( \frac{m_2^2}{m_1^2-m_2^2}\log\frac{m_2^2}{m_1^2} -\frac{m_3^2}{m_1^2-m_3^2}\log\frac{m_3^2}{m_1^2}\right)  & m_1 \neq m_2 \neq m_3 \\
\frac{1}{(m_1^2-m_2^2)^2}\left(m_1^2-m_2^2 +m_1^2\log\frac{m_2^2}{m_1^2}\right)  & m_2 = m_3 \\
\frac{1}{(m_1^2-m_3^2)^2}\left(-m_1^2+m_3^2-m_3^2\log\frac{m_3^2}{m_1^2}\right)  & m_1 = m_2 \\
\frac{1}{(m_1^2-m_2^2)^2}\left(-m_1^2+m_2^2-m_2^2\log\frac{m_2^2}{m_1^2}\right)  & m_1 = m_3 \\
-\frac{1}{2m_1^2}  & m_1 = m_2= m_3 \\
\end{array}
\right. ,
\end{align}
with $C_\mathrm{UV} = \frac{1}{\epsilon}-\gamma+\ln 4\pi $ being the UV divergent part. $C_{24}$ is expressed in terms of $C_0$ and $B_0$ as 
\begin{subequations} 
\begin{align}
C_{24} &= \frac{1}{4}+\frac{1}{2}m_1^2C_0 +\frac{1}{4}\left(B_0(2,3) +f_1 C_1+f_2 C_2\right) , \\
f_1 &= -m_1^2+m_2^2-p_1^2 ,\\
f_2 &= -m_2^2+m_3^2-(p_1+p_2)^2+p_1^2 ,\\
\left(
\begin{array}{l}
C_1  \\
C_2
\end{array}
\right) & = -X^{-1}
\left(
\begin{array}{l}
R_1  \\
R_2
\end{array}
\right) , \\
X &= \left(
\begin{array}{ll}
2p_1^2 & 2p_1\!\cdot\! p_2  \\
2p_1\!\cdot\!p_2 & 2p_2^2
\end{array}
\right) , \\
R_1 &= \left( f_1 C_0+B_0(3,1)-B_0(2,3)\right)  ,\\
R_2 &= \left( f_2 C_0+B_0(1,2)-B_0(3,1)\right)  ,\\
B_0(i,j) &= B_0(p_i,m_i^2,m_j^2) .
\end{align}
\end{subequations} 
%

%======================================%
%<<<<<<<<<<< bibliography >>>>>>>>>>>>>%
%======================================%
%%%%%%%%%%%%%%%%%%%%%%%%%%%%%%%%%%%%%%%%%%%%%%%%%%%%%%%%%%%%

%%%%%%%%%%%%%%%%%%%%%%%%%%%%%%%%%%%%%%%%%%%%%%%%%%%%%%%%%%%%

%%%%%%%%%%%%%%%%%%%%%%%%%%%%%%%%%%%%%%%%%%%%%%%%%%%%%%%%%%%%

\end{document}